\documentclass[conference,compsoc]{IEEEtran}

\usepackage{tikz}
\usetikzlibrary{plotmarks}
\usepackage{amsmath}

\usepackage[english]{babel}
\usepackage{graphicx}
\usepackage{url}
\usepackage[square, numbers]{natbib}
\usepackage{xcolor}
\usepackage{xspace}
\usepackage{booktabs}
\usepackage{tablefootnote}
\usepackage{algorithm}
\usepackage{algorithmic}
\usepackage{adjustbox}
\usepackage{comment}
\usepackage{subcaption}
\usepackage{enumitem}
\usepackage{multirow}

\usepackage{colortbl}
\usepackage{xintexpr, xinttools}

\graphicspath{{./graphics/}}

\expandafter\def\expandafter\UrlBreaks\expandafter{\UrlBreaks
  \do\a\do\b\do\c\do\d\do\e\do\f\do\g\do\h\do\i\do\j%
  \do\k\do\l\do\m\do\n\do\o\do\p\do\q\do\r\do\s\do\t%
  \do\u\do\v\do\w\do\x\do\y\do\z\do\A\do\B\do\C\do\D%
  \do\E\do\F\do\G\do\H\do\I\do\J\do\K\do\L\do\M\do\N%
  \do\O\do\P\do\Q\do\R\do\S\do\T\do\U\do\V\do\W\do\X%
  \do\Y\do\Z}

\providecommand{\UsePackageFor}[2]{ \ifx#2\undefined\usepackage{#1}\fi }

\UsePackageFor{amssymb}{\boxtimes}
\usepackage[normalem]{ulem}		
\usepackage{bbding}				
\usepackage{endnotes}			
\usepackage{hyperref}			
\usepackage{hyperendnotes}	
\usepackage{pdfcolfoot}			

\ifdefined\OrigFootnote\relax\else
	\newenvironment{FootnoteContent}{}{}
	\let\OrigFootnote\footnote
	\let\OrigFootnoteText\footnotetext
	\renewcommand{\footnotetext}[1]{\OrigFootnoteText{\begin{FootnoteContent}#1\end{FootnoteContent}}}
	\renewcommand{\footnote    }[1]{\OrigFootnote    {\begin{FootnoteContent}#1\end{FootnoteContent}}}
\fi

\definecolor{PurplePlum}{rgb}{0.1,0,0.55} 
\definecolor{Brown}{rgb}{0.5,.25,0}
\definecolor{Orange}{rgb}{1,.3,0}
\definecolor{Gray}{rgb}{.7,.7,.7}
\definecolor{DarkGreen}{rgb}{.1,.41,.1}

\newif\ifBleck
\newcommand\Bleck {\Blecktrue} 

\newcommand\Colour[1] {\color{#1}}

\newcommand\PrintToCLinks{	
  {\Colour{blue}\mbox{
    \hyperlink{w1619}{\sf$\rightarrow$~top}\quad
    \hyperlink{w1031}{\sf$\rightarrow$~toc}\quad
    \hyperlink{w1148}{\sf$\rightarrow$~lof}\quad
    \hyperlink{GreenRoom}{\sf$\rightarrow$~gr}\quad
    \hyperlink{EndNotes}{\sf$\rightarrow$~en}\quad
    \hyperlink{Sargasso}{\sf$\rightarrow$~sg}\quad
    \hyperlink{Index}{\sf$\rightarrow$~idx}
  }}
}
\makeatletter 
\newcommand\ToCLinks{
  \ifx\@onlypreamble\@notprerr		
    \hypertarget{w1619}{}			
  \else
    \AtBeginDocument{\hypertarget{w1619}{}}	
  \fi

  \ifBleck\else	
    \ifdefined\cofoot
      \cofoot{\PrintToCLinks}
      \cefoot{\PrintToCLinks}
    \else
      \def\@oddfoot{\PrintToCLinks}
      \def\@evenfoot{\PrintToCLinks}
    \fi
 \fi
}
\makeatother

\newif\ifEndNotes 

\newcommand\FnSym{{\scriptsize\PencilLeftDown\kern.1em}}		
\newcommand\EnSym {{$\bigtriangledown$}}

\def\MarkupsHowto{} 
\newcommand{\MarkupsHowtoAdd}[1]{\expandafter\def\expandafter\MarkupsHowto\expandafter{\MarkupsHowto{}#1}} 

\newif\ifMarkupsHowtoPrinted 
\newif\ifSuppress 

\newcommand\MakeMarkups[3][.]{

     \Suppressfalse
     \ifBleck\Suppresstrue\fi
     \ifx0#1\Suppresstrue\fi
     \ifx1#1\Suppressfalse\fi
     
     \expandafter\providecommand\csname#2x\endcsname {} 
     \ifSuppress\expandafter\renewcommand\csname#2x\endcsname{\relax}\else
                       \expandafter\renewcommand\csname#2x\endcsname{#3}\fi
                       
     \expandafter\providecommand\csname#2\endcsname {} 
     \ifSuppress\expandafter\renewcommand\csname#2\endcsname[1]{##1}\else
                       \expandafter\renewcommand\csname#2\endcsname[1]{{\csname#2x\endcsname##1}}\fi

     \expandafter\providecommand\csname#2d\endcsname {} 
     \ifSuppress\expandafter\renewcommand\csname#2d\endcsname[1]{\relax}\else
                       \expandafter\renewcommand\csname#2d\endcsname[1]{{\csname#2x\endcsname\sout{##1}}}\fi
                       
     \expandafter\providecommand\csname#2r\endcsname {} 
     \ifSuppress\expandafter\renewcommand\csname#2r\endcsname[2]{{##2}}\else
                       \expandafter\renewcommand\csname#2r\endcsname[2]{\csname#2d\endcsname{##1} \csname#2\endcsname{##2}}\fi

     \expandafter\providecommand\csname#2i\endcsname {} 
     \ifSuppress\expandafter\renewcommand\csname#2i\endcsname[1]{\relax}\else
                       \expandafter\renewcommand\csname#2i\endcsname[1]{\csname#2\endcsname{##1}}\fi

     \expandafter\providecommand\csname#2t\endcsname {} 
     \ifSuppress\expandafter\renewcommand\csname#2t\endcsname[1]{\relax}\else
                       \expandafter\renewcommand\csname#2t\endcsname[1]{{\csname#2x\endcsname{\mbox{$\langle\!\langle$}##1{\csname#2x\endcsname\mbox{$\rangle\!\rangle$}}}}}\fi 

     \expandafter\providecommand\csname#2b\endcsname {} 
     \ifSuppress\expandafter\renewcommand\csname#2b\endcsname[1][empty]{\relax}\else 
                       \expandafter\renewcommand\csname#2b\endcsname[1][\empty]{\ifx\empty##1\empty
                       	\label{#2-bookmark} 
                              \marginpar [\raggedleft\csname#2\endcsname{{\footnotesize\fbox{#2 working here}}~$\Longrightarrow$}]
                                                {\csname#2\endcsname{$\Longleftarrow$~{\footnotesize\fbox{#2 working here}}}}
                       \else 
                       	\marginpar [\raggedleft\csname#2\endcsname{\ifx\empty##1\empty\else\fbox{\tiny\parbox{8em}{\raggedright##1}}~\fi$\Longrightarrow$}]
                                                {\csname#2\endcsname{$\Longleftarrow$\ifx\empty##1\empty\else~{\tiny\fbox{\parbox{8em}{\raggedright##1}}}\fi}}\fi}\fi

     \expandafter\providecommand\csname#2TD\endcsname {} 
     \ifSuppress\expandafter\renewcommand\csname#2TD\endcsname{\relax}\else
                       \expandafter\renewcommand\csname#2TD\endcsname{\csname#2\endcsname{\fbox{#2 to do}}}\fi

     \expandafter\providecommand\csname#2Bar\endcsname {} 
     \ifSuppress\expandafter\renewcommand\csname#2Bar\endcsname{\relax}\else
                       \expandafter\renewcommand\csname#2Bar\endcsname{\csname#2\endcsname{\scriptsize\XSolidBrush}}\fi

     \expandafter\providecommand\csname#2f\endcsname {} 
     \ifSuppress\expandafter\renewcommand\csname#2f\endcsname[2][]{\relax}\else
      \expandafter\renewcommand\csname#2f\endcsname[2][\empty]{ 
        {\mbox{\csname#2x\endcsname\tiny$\boxtimes$}\marginpar{\hsize1cm\csname#2x\endcsname\fbox{\FnSym\footnotemark}}\relax 
        \footnotetext{\csname#2x\endcsname##2}}}\fi

     \expandafter\providecommand\csname#2e\endcsname {}
     \ifSuppress\expandafter\renewcommand\csname#2e\endcsname[1]{\relax}\else%
      \expandafter\renewcommand\csname#2e\endcsname[1]{%
       \global\EndNotestrue
       \mbox{\scriptsize\csname#2x\endcsname$\boxtimes$}\relax%
       \marginpar{\hsize1cm\csname#2x\endcsname\fbox{\EnSym\endnotemark%
                          \hypertarget{ENmark\thepage-\theendnote}{}~\hyperlink{ENtext\thepage-\theendnote}{{\Colour{blue}$\downarrow$}}}%
       }%
       {
        \def\zz{\noexpand#3}%
        \edef\z{~{[Endnote \theendnote\ %
        on p.\noexpand\hypertarget{ENtext\thepage-\theendnote}{}\thepage%
                    ~\noexpand\hyperlink{ENmark\thepage-\theendnote}{{\noexpand\Colour{blue}$\uparrow$}}]}%
        }%
        \expandafter\endnotetext\expandafter{\z\vspace{2ex}\\ ##1\newpage}%
       }
      }\fi

     \expandafter\providecommand\csname#2n\endcsname {}
     \ifSuppress\expandafter\renewcommand\csname#2n\endcsname[1]{\relax}\else%
      \expandafter\renewcommand\csname#2n\endcsname[1]{%
       \global\EndNotestrue
    \marginpar{{\tiny\endnotemark}\hypertarget{ENmark\thepage-\theendnote}{}~\hyperlink{ENtext\thepage-\theendnote}{}}
       {
        \def\zz{\noexpand#3}%
        \edef\z{~{\zz[Endnote (deferred) 
        from p.\noexpand\hypertarget{ENtext\thepage-\theendnote}{}\thepage%
        ]}%
        }%
        \expandafter\endnotetext\expandafter{\z\vspace{2ex}\\ ##1\newpage}%
       }
      }\fi

     \expandafter\providecommand\csname#2fe\endcsname {} 
     \ifSuppress\expandafter\renewcommand\csname#2fe\endcsname[2][]{\relax}\else 
      \expandafter\renewcommand\csname#2fe\endcsname[2][]{ 
       \def\File{##1}\relax
       \ifx\File\empty\csname#2f\endcsname{##2}\else 
        \global\EndNotestrue 
        \mbox{\scriptsize\csname#2x\endcsname$\boxtimes$}
        \marginpar{\csname#2x\endcsname\fbox{\FnSym\footnotemark}}\relax
        \footnotetext{~\csname#2x\endcsname##2\
                             --- See [\EnSym\endnotemark\hypertarget{ENmark\thepage-\theendnote}{}
                             \kern-.2em\hyperlink{ENtext\thepage-\theendnote}{{\Colour{blue}$\downarrow$}}].}\relax
       { 
         \def\zz{\noexpand#3}
         \edef\z{~{\zz[Endnote~\thefootnote~on~p.\noexpand\hypertarget{ENtext\thepage-\theendnote}{}\thepage
                     ~\noexpand\hyperlink{ENmark\thepage-\theendnote}
                     {{\noexpand\Colour{blue}\kern-0.1em$\uparrow$}]}}
                     {\noexpand\footnotesize\noexpand\newline\noexpand\hspace*{2em} (~from file {\noexpand\tt\File.tex}~)}
         }    
         \expandafter\endnotetext\expandafter{\z~\par\input{##1}\newpage}
        } 
       \fi 
      } 
     \fi 

     \ifSuppress\relax\else\ifBleck\relax\else
      \MarkupsHowtoAdd{\par\csname#2t\endcsname{
       $\backslash$\texttt{#2}$\cdots$\ markups are in \textbf{this} colour\ifx#1..\else\ifx1#1.\else, e.g.\ for #1.\fi\fi
       \ifMarkupsHowtoPrinted\relax\else 
        \global\MarkupsHowtoPrintedtrue 
        \begin{quote}\begin{tabular}{l@{\hspace{2em}}p{.7\linewidth}}
         \multicolumn{2}{l}{\texttt{$\backslash$MakeMarkups\ifx#1.\relax\else[#1]\fi\{#2\}\{{\it$\langle$colour command\/$\rangle$}\}}
         				 --- Defines the macros below:}\\
             & see comments at \texttt{$\backslash$MakeMarkups} definition. \\[1ex]
         \texttt{$\backslash$#2\{$\langle$text$\rangle$\}} & Sets \texttt{$\langle$text$\rangle$} in \texttt{#2}'s colour. \\
         \texttt{$\backslash$#2x} & Changes to \texttt{#2}'s colour (until end of context). \\
         \texttt{$\backslash$#2d\{$\langle$text$\rangle$\}} & Sets \texttt{$\langle$text$\rangle$} in \texttt{#2}'s colour with a strikethrough (i.e.\ delete). \\
         \texttt{$\backslash$#2r\{$\langle$this$\rangle$\}\{$\langle$that$\rangle$\}} &
          Strikes through \texttt{$\langle$this$\rangle$} and inserts \texttt{$\langle$that$\rangle$} (i.e.\ replace). \\
         \texttt{$\backslash$#2f\{$\langle$text$\rangle$\}} & Meta-comment: puts \texttt{$\langle$text$\rangle$} in a \texttt{#2}-footnote with a {\tiny$\boxtimes$} in the main text. \\
         \texttt{$\backslash$#2t\{$\langle$text$\rangle$\}} & Use for meta when  \texttt{$\backslash$#2f} isn't allowed (``Not in outer-par mode.'') \\
         \texttt{$\backslash$#2b[$\langle$optional$\rangle$]} & Marginal pointer, with label for hyper-linking directly there. \\
         \texttt{$\backslash$#2e\{$\langle$text$\rangle$\}} & Puts \texttt{$\langle$text$\rangle$} in a \texttt{#2}-endnote with a (big) $\boxtimes$ in the main text. \\[.5ex]
         \texttt{$\backslash$#2n\{$\langle$text$\rangle$\}} & Like \texttt{$\backslash$#2e}
         except there's no reference from the main text. Good for ``decluttering''
         when you still want to have the footnote- or endnote texts as reminders. \\[.5ex]
         \texttt{$\backslash$#2fe[$\langle$this$\rangle$]\{$\langle$that$\rangle$\}} & Makes a \texttt{$\backslash$#2f\{$\langle$that$\rangle$\}} that refers to a \\
           & \texttt{$\backslash$#2e\{$\langle$contents of file this.tex$\rangle$\}}. \\ 
           & Without the optional argument, acts as \texttt{$\backslash$#2f\{$\langle$that$\rangle$\}}. \\[.5ex]
         \texttt{$\backslash$#2Bar} & Inserts ``burn after reading'' symbol \csname#2Bar\endcsname, meaning
          \begin{quote}\begin{itemize}\setlength\itemsep{0pt}
           \item If yours is the only \csname#2Bar\endcsname\ in this (presumably someone else's) footnote, and you are happy that the footnote has been addressed,
           go ahead and comment-out the whole footnote. (The \csname#2Bar\endcsname\ is their request for you to ``approve and remove''.)
           \item If you are not happy, delete only your \csname#2Bar\endcsname\ and follow-on in the footnote
            (in your colour, i.e.\ with \texttt{$\backslash$#2x}) saying why you are not happy.
           \item If you are happy, but there are others' burn-after-reading symbols as well as yours, just delete yours; the other people have not yet responded.
          \end{itemize}
          \end{quote}
          The idea is that when everyone's happy, the last person will comment-out the meta-text. \\[0.5ex]
         \texttt{$\backslash$#2TD} & Inserts {\csname#2TD\endcsname}\ . \\
        \end{tabular}\end{quote}
       \fi
      }}
     \fi\fi
}

\newif\ifNoGreenRoom
\newcommand\MakeGreenRoom {\ifBleck\relax\else\ifNoGreenRoom\relax\else
\newcommand\NewGRLabel[1] {\OldGRLabel{GreenRoom-##1}} 
 \newcommand\NewGRRef[1] 
 {\expandafter\ifx\csname r@GreenRoom-##1\endcsname\relax\OldGRRef{##1}\else\OldGRRef{GreenRoom-##1}\fi}
 \let\OldGRLabel\label \let\label\NewGRLabel
 \let\OldGRRef\ref \let\ref\NewGRRef
 \hrule
 ~\\\begin{center}\Huge \hypertarget{GreenRoom}{Green Room}
 \end{center}~\\
 \hrule
\fi\fi}

\newcommand\EndGreenRoom  {\ifBleck\relax\else\ifNoGreenRoom\relax\else
\let\label\OldGRLabel
\let\ref\OldGRRef
\fi\fi}

\newif\ifNoEndNotes

\newif\ifNoSargasso
\newcommand\MakeSargasso {
 \hypertarget{Sargasso}{}
 \newcommand\NewLabel[1] {\OldLabel{Sargasso-##1}} 
 \newcommand\NewRef[1] 
 {\expandafter\ifx\csname r@Sargasso-##1\endcsname\relax\OldRef{##1}\else\OldRef{Sargasso-##1}\fi}
 \let\OldLabel\label \let\label\NewLabel
 \let\OldRef\ref \let\ref\NewRef
\ifBleck\end{document}\else\ifNoSargasso
\relax
\else
  \hrule
  ~\\\begin{center}\Huge Sargasso
  \end{center}~\\
  \hrule
 \fi\fi
}

\newcommand\EndSargasso  {\ifBleck\relax\else\ifNoSargasso\relax\else
\let\label\OldLabel
\let\ref\OldRef
\fi\fi}

\newcommand\EndDocument {\ifBleck\end{document}\fi} 

\newcommand\Cite[2][\empty] {{\Colour{red}\ifx#1\empty[#2]\else[#2,~#1]\fi}}

\Bleck		
\MakeMarkups[Sandra]{S}{\Colour{DarkGreen}}
\MakeMarkups[Ludovic]{L}{\Colour{purple}}
\MakeMarkups[Carmela]{C}{\Colour{blue}}
\MakeMarkups[Chris]{T}{\Colour{cyan}}

\newcommand{\ignore}[1]{}

\makeatletter
\def\customfootnote[#1]{\gdef\@thefnmark{#1}\@footnotetext}
\makeatother

\begin{document}
\date{}

\title{You get \texttt{PADDING}, everybody gets \texttt{PADDING}! You get privacy?\\
Evaluating practical QUIC website fingerprinting protections for the masses.}

\author{
  {\normalsize \rm Sandra Siby$^*$ ~~ Ludovic Barman$^*$ $^\ddag$ ~~ Christopher Wood$^\dag$ ~~Marwan Fayed$^\dag$ ~~Nick Sullivan$^\dag$ ~~ Carmela Troncoso$^*$}\\
\normalsize $^*$EPFL ~~ $^\dag$Cloudflare Inc.\\
} %

\def \numtraceroutes{974\xspace}
\def \numsites{150\xspace}
\def \sourceas{16276\xspace}
\def \asadvhigh{15169\xspace}
\def \asadvmed{16509\xspace}
\def \asadvlow{1229\xspace}

\def \mixed{\texttt{Mixed}\xspace}
\def \quicLarge{\texttt{LARGE}\xspace}
\def \quicM{\texttt{MAIN}\xspace}
\def \quicH{\texttt{HET}\xspace}
\def \quicSep{\texttt{TIME}\xspace}
\def \quicApr{\texttt{TIME2}\xspace}
\def \quicFirefox{\texttt{FIREFOX}\xspace}
\def \quicChrome{\texttt{CHROMIUM}\xspace}

\newcommand\eg{\emph{e.g.},\xspace}
\newcommand\ie{\emph{i.e.},\xspace}
\newcommand\etc{\emph{etc}.\xspace}
\newcommand\via{\emph{via}}
\providecommand{\etal}{\emph{et al.}\xspace}

\setlength\columnsep{15pt}
\newcommand{\para}[1]{\smallskip \noindent \textbf{#1}}
\newcommand{\parait}[1]{\smallskip \noindent \textit{#1}}
\newcommand{\ad}{\ensuremath{A}}
\newcommand{\obs}{\ensuremath{{\cal O}}\xspace}
\newcommand{\com}[1]{{\color{gray}#1}}


\maketitle
\customfootnote[$^\ddag$]{Ludovic Barman is currently at Google.}

\begin{abstract}
Website fingerprinting (WF) is a well-know threat to users' web privacy.
New internet standards, such as QUIC, include padding to support defenses against WF.
Previous work only analyzes the effectiveness of defenses when users are behind a VPN.
Yet, this is not how most users browse the Internet.
In this paper, we provide a comprehensive evaluation of QUIC-padding-based defenses against WF when users directly browse the web.
We confirm previous claims that network-layer padding cannot provide good protection against powerful adversaries capable of observing all traffic traces. 
We further demonstrate that such padding is ineffective even against adversaries with constraints on traffic visibility and processing power.
At the application layer, we show that defenses need to be deployed by both first and third parties, and that they can only thwart traffic analysis in limited situations.
We identify challenges to deploy effective WF defenses and provide recommendations to address them.
\end{abstract}

\section{Introduction}
\label{sec:introduction}
New standardization efforts have greatly increased the privacy of web traffic: \eg TLS Encrypted Client Hello (ECH)~\citep{ech} to encrypt Server Name Indication (SNI), or (Oblivious) DNS-over-HTTPS~\citep{doh, singanamalla2020oblivious} and DNS-over-TLS~\citep{dot} to encrypt DNS queries.
Yet, encryption alone cannot protect users' browsing history from traffic analysis.
Traffic-analysis attacks such as \emph{website fingerprinting} (WF), enable adversaries to infer which websites a user visits from the traffic patterns (\eg volume of packets exchanged or packets' size)~\citep{cheng1998traffic, hintz2002fingerprinting, herrmann2009website, miller2014know, siby2019encrypted, smith2021website}. 

QUIC is the next transport layer standard for the web that is seeing rapid adoption~\cite{quic_adoption}. 
In order to combat traffic analysis, the working group behind QUIC introduced a \texttt{PADDING} frame in the specification~\citep{quic_padding_frames}.
Recently, Smith \etal~\cite{smith2022qcsd} developed a client-side framework that can implement existing website fingerprinting defenses such as Tamaraw~\citep{cai2014tamaraw} and FRONT~\citep{gong2020zero} using the \texttt{PADDING} frame.
While they showed that it is possible to deploy website fingerprinting defenses solely from the client, their system is restricted to users in a VPN scenario where IP addresses cannot be used by an adversary to distinguish chaff traffic.
While a non-negligible amount of users browse the web through VPNs, this is not the case for the majority~\cite{VPNconsumer} who unfortunately will not benefit from such tools.

In this work, we investigate how to build defenses using the \texttt{PADDING} frame for \textit{the masses}, \ie outside of the VPN setting. 
Existing padding-based website-fingerprinting defenses differ on the layer they target and the information they use to inform the defense. There exist application-agnostic network-layer techniques that work independently of the website they are protecting~\cite{dyer2012peek, cai2014cs, cai2014tamaraw, juarez2016toward, gong2020zero}; 
application-layer-informed network-layer techniques that require prior knowledge (e.g., resources size, resources order, or total size) of the website trace they aim to protect to tailor the defense~\cite{rahman2020mockingbird, shan2021dolos};
and purely application-layer techniques that propose modifications directly on the resources~\cite{luo2011httpos, perry22rp, cherubin2017website}. 
We provide a comprehensive evaluation of both network- and application-layer defenses, to answer the question: can website-fingerprinting defenses be implemented solely at the network layer or does there need to be involvement of the application layer? This is motivated by the fact that QUIC is built in the user-space, thus allowing it to be more tightly-coupled with applications and making it a good candidate for application-layer-informed defenses. 
Moreover, in contrast to prior work, we emphasize \textit{practicality}, \ie we focus on how feasible the widespread deployment of these defenses are, so that normal users can be protected.

We assume a common scenario where websites are hosted behind content delivery networks (CDNs) and use privacy-preserving protocols such as TLS ECH. Users frequently encounter CDN-served resources while browsing -- as of November 2022, $\approx 44\%$ of the top million sites use CDNs~\cite{cdnstats}. In this scenario, the only available information to the adversary is the CDN server's IP address, and metadata such as the size of encrypted data, its timing, and its direction (sent by/to the server). We answer two questions:

\para{Can we build effective traffic analysis defenses at the network layer using the \texttt{PADDING} frame?} We study whether QUIC padding configured at the transport layer as envisioned by the standard can, on its own, protect users from \emph{website fingerprinting attacks}. We study adversaries with a wide range of capabilities: from powerful adversaries that can observe all communications between clients and servers, and have infinite storage and computation capability; to weaker, more realistic, scenarios in which the adversary can only observe partial traces and is restricted in its storage, computation, or bandwidth~\cite{nasr2017compressive}.
We find that:
\begin{itemize}[topsep=0pt,itemsep=-0.5pt,partopsep=1ex,parsep=1ex]
\item[\checkmark] Network-layer defenses which build on the QUIC \texttt{PADDING} frame to hide packets' sizes or inject dummy packets are not sufficient to prevent adversaries from inferring the websites users visit. Adversaries can use global trace information (\eg the total number of packets, or the total incoming size) to recognize websites ($>92\%$ F1-score). The adversary can successfully identify websites not just from their landing pages, but also from subpage visits, even if they have not previously encountered a visit to a specific subpage. These results hold even when the adversary uses limited information from typical network statistics (\eg NetFlow). Only when the adversary  observes a very small percentage of the page does the adversary's performance decrease to close to random guessing. 

\item[\checkmark] The centralization of web resources on the Internet, in particular, in the hands of Google, creates a favorable setting for the adversary. Traffic analysis on solely the timing of Google resources fetched by a web page achieves $>77\%$ F1 score, requiring \emph{four orders of magnitude less data than using full traces}. Moreover, the surface of attack is increased from only ASes between the client and the first-party domain host to any AS between the client and Google.
\end{itemize}
 
\para{Can application-layer defenses effectively thwart traffic analysis?} We reveal that network-layer defenses based on padding QUIC traffic are ineffective because they cannot efficiently hide the most important feature for the adversary: the total size of a website. This is because this information is not known at the network layer. We, explore whether this information can be effectively collected at the application layer, and whether it can be used to inform padding algorithms in an effective manner:
We find that:
\begin{itemize}[topsep=0pt,itemsep=-0.5pt,partopsep=1ex,parsep=1ex]
\item[\checkmark] Websites contain a vast amount of resources. These resources are served, in a significant number of cases, by third parties (33\% in our dataset). We demonstrate that if all parties do not participate in defending against WF, an adversary can successfully identify pages ($>91\%$ F1-score), leaving users vulnerable.

\item[\checkmark] We identify current web development practices that hinder the deployment of effective website fingerprinting defenses if users are not behind a VPN. We show that, unless those practices do not change, protecting users is close to impossible as it would require significant bandwidth overhead and coordination among parties under very dynamic conditions. We provide recommendations to guide future efforts and pave the way for the existence of effective defenses against website fingerprinting attacks.
\end{itemize}

\para{Ethical considerations:}
We conduct traffic-analysis attacks against a deployed technology (QUIC).
We do not perform any collection or analysis of real users' traffic. 
We only collect our own traffic, generated by an automated browser. 
We uncover vulnerabilities in the proposed defenses, which would put at risk, network users, if deployed.
We believe that the benefits of our research, which can guide current and future standardization efforts, outweigh these risks, by avoiding deployments that could give users a false sense of security.
We have performed responsible disclosure of our findings to QUIC's IETF WG.\@

\section{Background \& Related Work}
\label{sec:background}
\para{QUIC.} QUIC is a connection-oriented protocol built on top of UDP that aims to provide low-latency, multiplexed, secure communication with less head-of-line blocking and faster connection migration~\citep{quic_rfc}.
QUIC was standardized in May 2021 and is currently being developed by the IETF.\@
QUIC is the transport protocol for HTTP/3. 
Adoption of QUIC and HTTP/3 has been rising (as of November 2022, they are used by 25.5\% of the top 10 million websites~\citep{quic_adoption}).
Of particular relevance for our work is the QUIC \texttt{PADDING} Frame. 
The IETF QUIC draft describes it as a frame with no semantic value, that can be used to increase packets size and to provide protection against traffic analysis~\citep{quic_padding_frames}.

\para{Website fingerprinting attacks.}
In website fingerprinting, an adversary analyzes network traffic to infer the website visited by a user. 
The adversary builds a classifier trained on features obtained from website network traces. 
These features can be selected manually or via automatic extraction.

The most relevant attacks that rely on manual features are Wang et al.'s k-Nearest Neighbors (k-NN) classifier based on 4226 manually-selected features~\citep{wang2014effective}; Panchenko et al.'s Support Vector Machines (SVMs)-based classifier, based on cumulative sums of packet lengths~\citep{panchenko2016website}; and Hayes and Danezis~\citep{hayes2016kfingerprinting} \textit{k}-fingerprinting method (\textit{k}-FP), which models web fingerprints as the leaves of a random forest built on 150 manually-selected features.

On the automatic extraction side, Rimmer et al.~\citep{rimmer2018automated} use deep learning neural networks (DNNs) to produce attacks that perform as well as manual approaches. Sirinam et al.~\citep{sirinam2018deep} build on Rimmer et al. to develop an attack that achieves high accuracy, even in the presence of defended traces.
Last, Bhat et al.~\citep{bhat2018var} propose Var-CNN, a hybrid strategy that achieves high accuracy with deep learning even in the presence of limited data.
It does so by relying on ResNets trained on packet directions, packet times, and manually extracted summary statistics.

\parait{Website fingerprinting on QUIC traffic.} 
Smith et al.~\citep{smith2021website} study the impact of co-existence of TCP and QUIC on the performance of website fingerprinting using \textit{k}-FP and Var-CNN.\@
They conclude that, while QUIC traffic is not difficult to fingerprint, classifiers trained on TCP traffic do not perform well on QUIC traffic, and that jointly classifying both protocols is hard.
To enable comparison with the state-of-the-art, we also use \textit{k}-FP and Var-CNN in our evaluation.

\para{Website fingerprinting defenses.}
Dyer et al.~\citep{dyer2012peek} show that \emph{network-layer} padding- and morphing-based countermeasures are ineffective in thwarting traffic analysis because they fail to hide coarse packet features. They propose Buffered Fixed-Length Obfuscation (BuFLO), which pads packets to a fixed size and sends them at intervals of time. BuFLO results in a huge overhead. CS-BuFLO~\citep{cai2014cs} and Tamaraw~\citep{cai2014tamaraw} are more efficient, but still impractical.
Works such as WTF-PAD~\citep{juarez2016toward} and FRONT~\citep{gong2020zero} provide a better trade-off by injecting dummies at appropriate positions in a trace.
WTF-PAD injects dummy packets using pre-defined distributions of inter-arrival times to detect gaps, and FRONT injects dummy packets in the front portion of traces, which is known to contain the most information for fingerprinting. Both approaches achieve low protection against deep-learning-based attacks~\citep{sirinam2018deep, bhat2018var, shan2021dolos}.

Other defenses employ \emph{adversarial perturbations} to cause deep-learning-based classifiers to misclassify traces. 
These defenses incur lower overhead than prior work.
Mockingbird~\citep{rahman2020mockingbird} applies perturbations to convert traces into a target trace.
It converts traces into bursts (where a burst is a set of contiguous packets in one direction), and perturbs these bursts rather than the raw trace.
To compute the perturbation, Mockingbird requires the defender to know the entire trace in advance, which is infeasible in practice.
Nasr et al.~\citep{nasr2021blind} tackle this issue with Blind, a defense that pre-computes blind perturbations that can be applied to live network traffic.
Shan \etal~ \citep{shan2021dolos} show that ~\cite{nasr2021blind} offers lower protection when the adversary trains on perturbed traces. 
They propose Dolos, which applies adversarial patches or pre-computed sequences of dummy packets to protect network traces.
Dolos utilizes a user-side secret to generate patches, making it hard for an adversary to generate the same patch, thereby reducing the risk of adversarial training.

Mockingbird, Dolos, and Blind provide protection to Tor traffic.
Tor cells have constant size, as opposed to QUIC packets.
These defenses do not account for packet sizes, hindering their adoption to protect QUIC traffic.
Mockingbird~\citep{rahman2020mockingbird} only considers packet directionality when building dummy bursts.
It is unclear how to adapt it to QUIC traffic; a burst can correspond to many QUIC packet sequences.
Blind~\citep{nasr2021blind} does not use packet size information in their website fingerprinting evaluation.
The paper contains a size perturbation technique (tailored to Tor) and uses it for their flow correlation experiments. 
However, even after communicating with the authors, we were unable to find where in their implementation~\cite{blanket} one can configure this technique, nor are there details about how to configure it for non-Tor traffic.
Dolos~\citep{shan2021dolos} uses solely direction features to compute patches.
Adapting it to QUIC would require integrating size information, for which there is no place in their implementation, with no guarantee that the patches would still be effective. 
Moreover, Dolos requires a prior connection to the website in order to pre-compute patches which can be used in future connections.  
Hence, it is predicated on information from the application layer, i.e, which website is being visited. This is in line with our findings on the importance of having the application layer informing any network layer defense (Section~\ref{sec:applayer}). 
Due to the challenges associated with adapting these Tor-tailored systems to our QUIC scenario, in our work, we compare to FRONT~\citep{gong2020zero}, which Smith et al. show it can be implemented for QUIC~\cite{smith2022qcsd}. 

Finally, there are defenses at the \emph{application layer}. 
Luo et al.~\citep{luo2011httpos} propose HTTP Obfuscation (HTTPOS), a client-side defense that modifies features on the TCP and HTTP layers and uses HTTP pipelining to obfuscate HTTP outgoing requests. Randomized Pipelining~\citep{perry22rp} improves this defense by randomizing the order of the HTTP requests queued in the pipeline. 
Subsequent works have shown HTTPOS and Randomized Pipelining to be ineffective against traffic analysis attacks~\citep{cai2012touching, wang2014effective}.
Cherubin et al.~\citep{cherubin2017website} developed client- and server-side defenses, LLaMA and ALPaCA respectively, tailored towards Tor onion services. These defenses only work well in scenarios with low third party content prevalence, lack of dynamic page content, and JavaScript disabled. 
In our work, we propose defenses inspired by ALPaCa, and study their performance in dynamic web scenarios where there are a large amount of third party resources (see~\autoref{sec:applayer}).

\section{Adversarial Model and Datasets}
\label{sec:model}
We assume a local passive eavesdropper \ad{} located at some vantage point between an honest client and an honest Web host.
\ad{} observes all network traffic passing through this vantage point and records some portion of it.
The goal of the adversary is to infer the domain visited by the user.

The adversary \ad{} observes IP packets.
They do not possess any decryption keys, and rely only on the size and timing of the observed packets.
We assume that DNS queries are done in a private manner (\eg via DoH~\citep{doh} with appropriate padding~\citep{siby2019encrypted}) and reveal no information to \ad{}.
\ad{} focuses on Web traffic, and filters out packets that are not TLS or QUIC packets.
Using the appropriate fields in the headers (IP addresses, ports, QUIC connection IDs), \ad{} is able to identify packets that belong to the same connection.

We call \ad{}'s \emph{observation} a collection of flows corresponding to the network connections generated when the user browses a single website.
Each flow contains $[\text{IP}_\text{source}, \text{IP}_\text{dest}, p_1, p_2, \dots]$, where packets $p_i$ are (time,size)-tuples $(t_i, \pm s_i)$.
Negative sizes indicate packets from the server to the client, and positive sizes indicate packets in the opposite direction. 

\begin{figure}
	\centering
	\small
	\includegraphics[width=\linewidth]{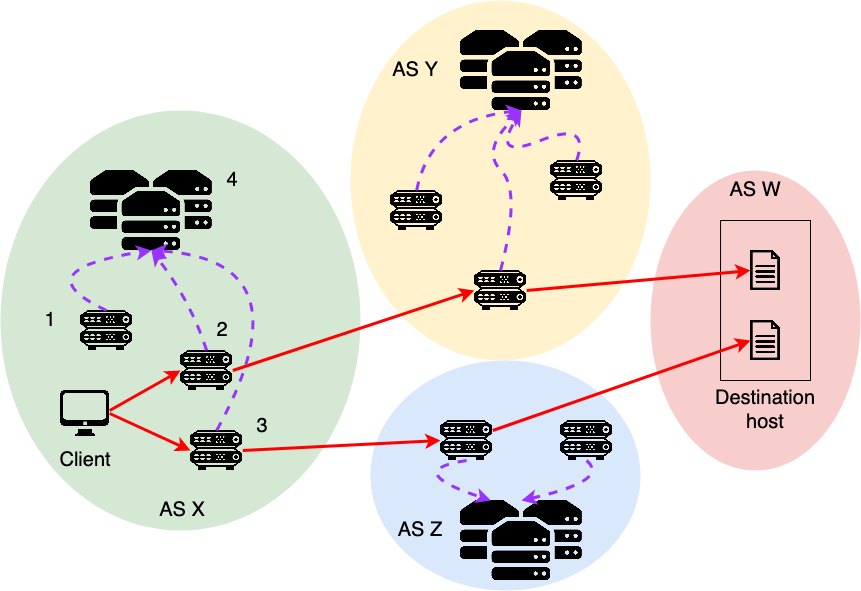}
	\caption{An adversary can be on any AS (X, Y, Z, or W) with vantage points on the client's traffic path (solid red arrows).
	The vantage points transmit recorded data to a location that can perform traffic-analysis at scale (dotted purple arrows). If the adversary is AS X, vantage points 2 and 3 transmit recorded traffic to location 4.}
	\label{fig:model}
	\vspace{-0.4cm}
\end{figure}

\para{Vantage Points.} Following prior work~\citep{feamster2004location, murdoch2007sampled, johnson2013users} we consider each AS on the path of the client traffic as a realistic adversary. 
Each AS' middlebox, router, or switch that routes traffic from a client is a potential vantage point for the adversary to collect this client's traffic.
In \autoref{fig:model}, we depict a client located in AS $X$ accessing two webpages hosted on an IP in AS $W$.
The client traffic is represented by red lines.
If the adversary controls AS $X$, they observe all the traffic related to the page visits. This is the adversary typically considered in the website fingerprinting literature~\citep{cheng1998traffic, hintz2002fingerprinting, herrmann2009website, miller2014know, wang2014effective, panchenko2016website, hayes2016kfingerprinting, rimmer2018automated, sirinam2018deep, bhat2018var, smith2021website}.
If the adversary controls AS $Y$ or AS $Z$, however, they would have limited visibility on the traffic, \ie they might not observe traffic from \emph{all} clients' visited webpages, or for each observed web page, they might only observe a portion of the traffic (\eg the loading of some resources).
We note that it is possible for an adversary to control multiple ASes or an IXP (where traffic from multiple ASes can traverse)~\citep{murdoch2007sampled, johnson2013users}.

\subsection{Website fingerprinting}
As in previous work, we implement website fingerprinting attacks as a supervised learning problem.
The adversary identifies the IP that contains the domains that they want to target.
Then, the adversary enumerates all the domains on that IP, and collects web traffic traces from these domains.
The adversary extracts features from these traces, and uses the feature vectors to train a classifier.
Given a new trace, this classifier predicts which website it belongs to.

We implement the attack using a random forest classifier, a simple and effective model frequently used in website fingerprinting; and Var-CNN~\citep{bhat2018var} to validate our results against the state of the art~\citep{smith2021website}.
We use the set of features proposed by Hayes \etal~\citep{hayes2016kfingerprinting} for performing website fingerprinting on Tor since this is a comprehensive set of features from previous related works.
To adapt them to the QUIC case, we add features about packet size frequencies.
Since QUIC's maximum payload size (1400 B) is smaller than that of TLS (16 kB), we compute frequencies of packet sizes up to 16 kB to encompass both QUIC and TLS traffic.

\subsection{Adversarial Capabilities and Goals}

We assume applications and websites hosted behind content delivery networks that use QUIC, and protocols to protect sensitive information, e.g., TLS ECH. 
The information available to the adversary comprises the server's IP address, which is determined by the content delivery network, and any application-specific metadata such as the size of encrypted data, which is determined by the application. 
As the IP is known, an adversary can enumerate the domains this IP hosts (\eg through DNS reverse lookups or from DNS scans using public name sources, including Certificate Transparency logs), collect pages associated with the domain, and train a classifier on these samples.
This is a more tractable scenario than that of the open world Internet. 

We study two adversarial models: unconstrained and constrained, depending on the adversary's visibility on traffic and the resources they can dedicate to fingerprinting. We also study different adversarial goals and client setups that may impact the adversary's success. We summarize details of the datasets we collect in \autoref{table:datasets}.

\begin{table}
	\caption{Overview of datasets. All datasets except \quicChrome are collected using Firefox.}
	\centering
	\resizebox*{\columnwidth}{!}{%
	\begin{tabular}{lccc}
		\toprule
		\textbf{Experiment} & \textbf{Identifier} & \textbf{\# webpages} & \textbf{\# samples} \\
		\midrule
		Landing pages main set (Mar'21) & \quicM & 150 & 40 \\
		Landing pages large set (Nov'22) & \quicLarge & 350 & 40 \\
		Influence of time (Sept'21) & \quicSep & 145 & 40 \\
		Influence of client (May'22) & \quicFirefox & 131 & 40 \\
		Influence of client (May'22) & \quicChrome & 131 & 40 \\
		Domains (Nov'22) & \quicH &  60 & 35 \\
		\bottomrule
	\end{tabular}
	}
	\label{table:datasets}
\end{table}

\subsubsection{Unconstrained adversary}

An \emph{unconstrained} adversary can observe and process \emph{all} the traffic associated with a web page visit. We assume this adversary can have one of two goals:

\para{Fingerprinting landing pages (Homogeneous Closed World).} 
In this scenario, the most common in the literature, we assume that users only visit landing pages~\cite{herrmann2009website, hayes2016kfingerprinting, rimmer2018automated, sirinam2018deep, bhat2018var}. 
Thus, the adversary only needs to collect landing pages to train their model. 
The training and testing sets contain the same web pages, for which the traces vary due to the page content changing when traces are collected at different points in time. 
As QUIC leaves IPs unprotected, the adversary can limit the anonymity set of a web page, and this corresponds to a classic \emph{closed world} classification. 

\parait{Dataset creation.} 
Prior works rely on lists of most visited sites (Alexa, Umbrella, \etc) to build datasets.
These domains are typically hosted on different IPs, and thus would not be in the same anonymity set in our adversarial model.
This would only happen if the client uses a VPN, which is out of the scope of this work.
We work with a CDN provider to identify realistic anonymity sets that could be targeted by the adversary. 
We identify $13,744,979$ unique domain names on the CDN in Mar 2021, and use \texttt{zdns}~\cite{izhikevich2022zdns} to find that these domains are hosted on $593,338$ IPs or anonymity sets for the adversary.

The anonymity set sizes follow the same distribution as that found by Patil \etal~\citep{patil2019can} (see \autoref{fig:distribution-clusters} in Appendix).
We find that $60\%$ of these domains are hosted on a unique IP address: when observing one of these IPs, the adversary is certain of which domain is being visited.
Therefore, these domains are out of scope for our study.
Only $50k$ IPs ($8.5\%$ of the dataset) host more than $150$ domains, with one hosting as many as $56,319$ domains.
We choose a cluster of $150$ websites for our experiments: which is a hard scenario among the $91.5\%$ of the IPs served by the CDN provider. 
This cluster has a high percentage of TLS traffic: only $4\%$ (std $1.7\%$) of the traffic is transmitted over QUIC.
This results in the classifier focusing on TLS-specific features, preventing us from drawing meaningful conclusions about QUIC traffic's vulnerability or QUIC-oriented defenses (results in Appendix~\ref{app:feature_importance}).
We also run experiments with other clusters of similar size and obtain comparable results.

To address this problem, following the example of Smith \etal~\citep{smith2021website}, we crawl Alexa 1M~\citep{alexa1M}, Umbrella 1M~\citep{umbrella1m}, and Majestic 1M~\citep{majestic1M}.
We perform HAR captures\footnote{A copy of the ``Network'' tab of the Firefox Developer Tools console.} and we identify the protocols used by those websites.
Unlike~\citep{smith2021website}, we consider only domains that primarily use QUIC. We select 150 of these domains, and collect a dataset of traces: \quicM (collected in March 2021).
\quicM has $70\%$ of all traffic over QUIC (std $3\%$). 
We also collect a larger dataset of 350 domains, \quicLarge (collected in November 2022, with $50\%$ QUIC traffic), to evaluate whether fingerprinting performance changes with an increase in the dataset size. 
We note that in given our measurements, a larger dataset would be rare in a CDN scenario-- only 2.2\% of the IPs in our CDN dataset host more than 350 domains.

\para{Fingerprinting domains (Heterogeneous Closed World).} 
The previous scenario is somehow artificial in the sense that users visit more than the landing pages of domains. 
Thus, only training on landing pages would not work well in a realistic deployment.
To model a domain, the adversary needs to train on both landing and subpages.
Yet, due to the visited IPs being visible, the classification problem is still a closed world: the adversary has a finite set of domains to associate traffic traces to.
We collect a dataset to evaluate how the adversary performs when also training on heterogeneous subpages of a domain. 

\parait{Dataset creation.} 
We collect \quicH, a dataset consisting of subpages for each domain hosted by the IP, instead of multiple samples of the landing page. 
We enumerate all pages that can be visited from the landing page (sharing the same domain as the landing page), and the pages that can further be visited from those subpages.
Since the classifiers require a reasonable number of training samples, we limit our study to a set of 60 domains hosted by the target IP which have at least 35 subpages. 

\para{Website-fingerprinting robustness} 
We collect data to study additional factors that may influence the performance of the adversary.
To study the stability of website fingerprinting attacks over time, we collect a QUIC-dominated set of traces from the same domains as in \quicM in September 2021 (\quicSep).
In order to evaluate the influence of the client's browser, we collect \quicFirefox and \quicChrome -- two datasets collected during the same time period, with the Firefox and Chromium browsers.

\para{Data collection.}
To build the datasets, we collect PCAP network traces from visits to each landing page of the domains in our list.
We use Firefox isolated in its own network namespace (using \texttt{netns}), enabling HTTP3, and disabling telemetry and auto-update settings to minimize extraneous traffic.
We record $40$ samples for each website.
For each sample, we clean the caches by creating a fresh Firefox profile. 
We extract well-formed TLS and QUIC packets from the traces.
To avoid relying on plaintext markers, we follow the approach of Smith \etal~\citep{smith2021website}, and only extract the time and size of the sent and received packets.
We use Firefox 88.0 for \quicM and \quicSep, Firefox 98.0 for \quicLarge, \quicFirefox and \quicH, Chromium 101.0 for \quicChrome.

\subsubsection{Constrained adversary}
We assume that vantage points do \emph{not} have the capability to run machine-learning tasks~\citep{bai2020fastfe, barradas2021flowlens}.
They must mirror (part of) the traffic to a suitable location for analysis (purple dotted arrows in~\autoref{fig:model}).
This location processes the traffic: it extracts features and performs classification to identify the page visited by the client.
In practice, mirroring all traffic is prohibitively expensive~\citep{nasr2017compressive}.
Thus, we also study a \emph{constrained} adversary that only transmits summaries of locally computed statistics from sampled data~\citep{tammaro2012exploiting, carela2011analysis}. 

We measure the adversary's cost to perform the attack in terms of the bandwidth they require to collect and process the traffic traces. 
Bandwidth is a proxy for the required storage, as the adversary needs to store the transmitted information to query the machine learning model and possibly to retrain it.
The computational cost is also proportional to the bandwidth, as the number and cost of operations needed to extract features depend on the length of the traces transmitted.
We evaluate the adversary's success using filtered versions of the datasets described above.

\section{Network-layer \texttt{PADDING}-based defenses}
\label{sec:powadv}

In this section, we study whether \texttt{PADDING}-based network-layer defenses can thwart website fingerprinting. 

\subsection{Unconstrained Adversary}

First, we evaluate the effectiveness of defenses against a powerful adversary that can observe \emph{all} the traffic associated with a site visit, and can store and process \emph{all} the traffic that it observes. 
Such an adversary could be AS $X$ in \autoref{fig:model}, if this AS would not have bandwidth or storage constraints.  

\subsubsection{Unprotected Traces}

First, we determine the performance of the adversary on unprotected QUIC traces on which no \texttt{PADDING} frames are used. We study different scenarios according to the goals defined in Sect.~\ref{sec:model}.

\para{Fingerprinting landing pages (Homogeneous closed world).} A random forest classifier obtains a F1-score of $95.8\%$ when users only browse landing pages (\quicM, Table~\ref{tab:perf-ip-wf}, first row). 
The results are orders of magnitude better than random guessing ($0.67\%$). We repeated this experiment with ten different 150-sites sets and obtained similar results.

These results hold when considering a large anonymity set for the CDN scenario (\quicLarge, second row). 
We also try Var-CNN~\citep{bhat2018var} on \quicM and we obtain an F1-score of 92.28\%. 
Since the performance of random forest is better and they give us the advantage of interpretability, we use random forest for all our experiments. As we do not observe a large difference between \quicM and \quicLarge, we pick \quicM (which shows a greater advantage for the adversary) and datasets of equivalent sizes for our remaining experiments.

\begin{table}
	\caption{Performance of website and IP fingerprinting on the \quicLarge and \quicM datasets (10 experiments).}
	\centering
	\begin{tabular}{l r}
		\toprule
		\textbf{Method} & \textbf{F1-score}\\
		\midrule
		Website fingerprinting (\quicM) & 95.8 (std. dev. 0.4)\\
		Website fingerprinting (\quicLarge) & 93.7 (std. dev. 0.2)\\
		IP fingerprinting with primary IP (\quicLarge) & 70.6 (std. dev. 0.1)\\
		IP fingerprinting without primary IP (\quicLarge) & 37.5 (std. dev. 0.1)\\
		\bottomrule
	\end{tabular}
	\label{tab:perf-ip-wf}
\end{table}

Unlike in prior work~\citep{smith2021website} which considered a VPN scenario, in our threat model, the adversary can observe the IP addresses of the communication end-points. 
Hoang \etal~\cite{hoang2021ipwf} showed that destination IPs can be used to identify the websites visited by a user.
To understand whether the adversary can take advantage of the IPs, we conduct the IP fingerprinting process described in ~\cite{hoang2021ipwf}.
We perform repeated DNS resolutions of the resources loaded by each website in \quicLarge to build a database of IP fingerprints.
For each website, we perform 20 resolutions per day over a period of 10 days.
We perform these resolutions right after the trace collection of \quicLarge, so that IP-domain mappings are as close as possible to our traces.

We use the basic-IP fingerprinting methodology described in~\citep{hoang2021ipwf} to find the best match between our \quicLarge traces and the fingerprints.
In this method, first the adversary matches the IPs of the primary domain to the fingerprints to get a set of candidate sites. 
Then, they match the IPs of secondary domains against the candidates' fingerprints to obtain a final match.
We perform two experiments: in the first, we perform the IP fingerprinting as described.
In the second, to match our threat model in which all sites are hosted on a single IP by a CDN, we perform the matching solely on secondary IPs.
The results for the two experiments are shown in the third and fourth rows of Table~\ref{tab:perf-ip-wf}.

We see that IP fingerprinting is less accurate than website fingerprinting, especially when the primary IP is not available.
When the primary IP is used, IP fingerprinting presents a somewhat bi-modal behavior. It is very reliable for $32\%$ of the sites for which it correctly classifies \emph{all} the traces (vs. $18\%$ fully correct classification by website fingerprinting). But, it performs very poorly for the $68\%$ remaining. 
Website fingerprinting is not perfectly reliable for more websites, but overall performs better.

The advantage of website fingerprinting grows when the primary IP is unavailable.
Then, website fingerprinting correctly predicts all samples for $\approx$ 35\% of sites as compared to $\approx$14\% for IP fingerprinting.
In other words, IP fingerprinting is useful when there is a stable primary IP to provide a strong signal.
Given our threat model, in the remaining sections, we will only use website fingerprinting as we are studying the effect of website fingerprinting defenses.

\para{Fingerprinting domains (Heterogeneous closed world).} 
Next, we study the case in which users visit any website within a domain. 
We use the \quicFirefox (as the fingerprinting landing webpage baseline) and \quicH datasets.
These datasets use the same browser and are collected during the same time period.
We use only the domains found in \quicH when attacking the \quicFirefox dataset. 
Table~\ref{tab:heterogeneous} shows the results.

For \quicH, we evaluate two cases. 
In the first case (second row), the adversary has trained on all the subpages of a site, i.e., the training set has the same distribution as the test set.
In the second case (third row), the adversary has trained on a subset of subpages and has to classify unseen subpages, i.e., there is a shift between the training and testing sets distribution.
We use 30 subpage samples per domain to train the classifier, and 5 subpage samples to test (with different train and test samples per fold).
We conclude that, even when the adversary has to classify unknown subpages, there is a small performance drop, maximum $\approx$ 3\%.
The adversary's success is most likely caused by subpages of the same site sharing many resources, leading to very similar traces. 

\begin{table}[t!]
  \centering
	\caption{Fingerprinting Landing webpages vs. Website.}
  \resizebox*{\columnwidth}{!}{%
	\begin{tabular}[c]{ l  r  r }
	\toprule
			\textbf{Scenario}	& \textbf{Dataset}  & \textbf{F1-score}  \\
		\midrule
			Landing webpage (baseline) & \quicFirefox  & 97.8 (std. dev. 0.4) \\
			Website (all subpages known) & \quicH  & 96.4 (std. dev. 0.9)  \\
			Website (unknown subpages) & \quicH  & 94.1 (std. dev. 1.1) \\
		\bottomrule
	\end{tabular}
	}
  \label{tab:heterogeneous}
\end{table}

\subsubsection{Robustness}

Next, we study factors that can influence the adversary's fingerprinting performance.

\parait{Influence of time.} The results in \autoref{tab:perf-ip-wf} assume that the adversary collected the training set for their classifier close in time to their attack. 
We study the performance of the attack when trained on traces collected at a time different than testing in \autoref{tab:time}.
We use the \quicM and \quicSep datasets, collected 6 months apart. 
We note that \quicSep has 145 domains instead of 150, due to failures in data collection (unreachable sites and repeated timeouts), hence we consider only these domains. 
For both datasets, the classifier performance remains consistent when it is trained on data collected close to the attack.
However, the performance drops significantly when used on data collected at a different point in time.
This indicates that for an adversary to be successful, they would need to update their training data with more recent samples to keep up with constantly evolving web pages.

\begin{table}[t!]
  \centering
	\caption{Influence of time. F1-score when training on the row dataset and testing on the column dataset.}
	\begin{tabular}[c]{ l  r  r  }
	\toprule
			\textbf{F1-score}	& \textbf{\quicM}  & \textbf{\quicSep} \\
		\midrule
			\textbf{\quicM} & 95.8 (std.\ dev. 0.4) & 35.2 (std. dev. 1.2)  \\
			\textbf{\quicSep} & 17.5 (std. dev. 2.1)  & 96.4  (std. dev. 0.2) \\
		\bottomrule
	\end{tabular}
  \label{tab:time}
\end{table}

\parait{Influence of client.} 
We consider the influence of the client setup on the adversary's performance.
We use two datasets, \quicFirefox (Firefox 98.0) and \quicChrome (Chromium 101.0), collected at the same time.
Both datasets consist of 131 domains after accounting for collection failures.
Table \autoref{tab:client} shows that the classifier performs as expected when trained and tested on the same client setup.
Similar to previous work ~\cite{siby2019encrypted}, we see that when the setup changes, performance drops, indicating that an adversary would need a classifier tailored to the client setup.  
A reason may be that Firefox traces are generally longer, possibly due to activities such as contacting OCSP servers.

 \begin{table}[t!]
  \centering
	\caption{Influence of client. F1-score when training on the row dataset  and testing on the column dataset.}
	\begin{tabular}[c]{ l  r  r }
	\toprule
			\textbf{F1-score}	& \textbf{\quicFirefox}  & \textbf{\quicChrome}  \\
		\midrule
			\textbf{\quicFirefox} & 95.6 (std. dev. 0.3) & 35.6 (std. dev. 2.4) \\
			\textbf{\quicChrome} & 22.9 (std. dev. 1.6) &  92.9 (std. dev. 0.3)\\
		\bottomrule
	\end{tabular}
  \label{tab:client}
\end{table}

\subsubsection{\texttt{PADDING}-based network defenses}

To study the effectiveness of defenses based on the QUIC \texttt{PADDING} frame, we focus on the worst case: fingerprinting undefended landing pages visited with the same client using fresh training datasets. We run our experiments on the \quicM dataset.

We explore defense strategies that hide local and global features, assuming the presence of an implementation that perfectly protects these features.
\autoref{table:quic-defenses-full-view} reports the results against difference defenses (first row are undefended traces).

\begin{figure}
	\centering
	\includegraphics[width=\linewidth]{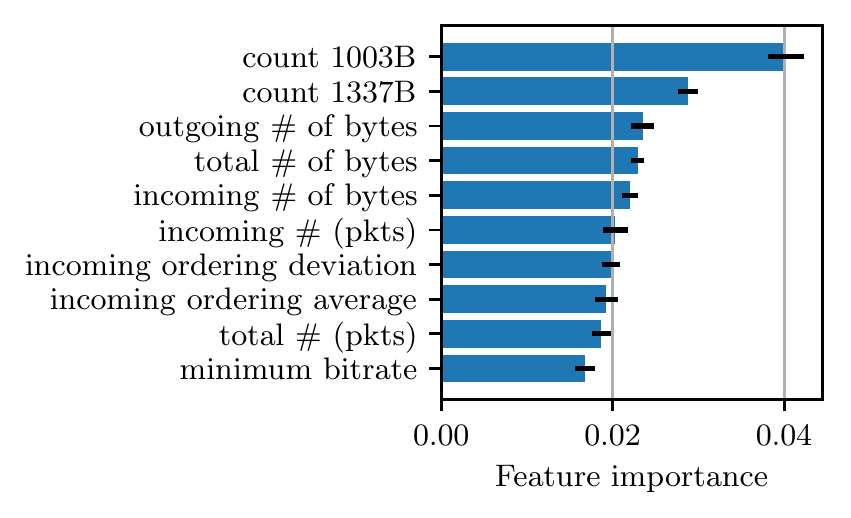}
	\vspace{-0.6cm}
	\caption{Feature importance for \quicM\ }
	\label{fig:quic-150-fi}
	\vspace{-0.3cm}
\end{figure}

\parait{Hiding local features.} 
The top predictive features for \quicM are all size-based (\autoref{fig:quic-150-fi}).
When the defense hides individual packet sizes, the attacker performance slightly decreases with respect to undefended traces.
Padding individual packets poorly hides the total transmitted volumes, which becomes the top feature once individual sizes are removed. 
Hiding timings also has minimal impact on the reduces the adversary's performance.

\begin{figure}
	\centering
	\includegraphics[width=\linewidth]{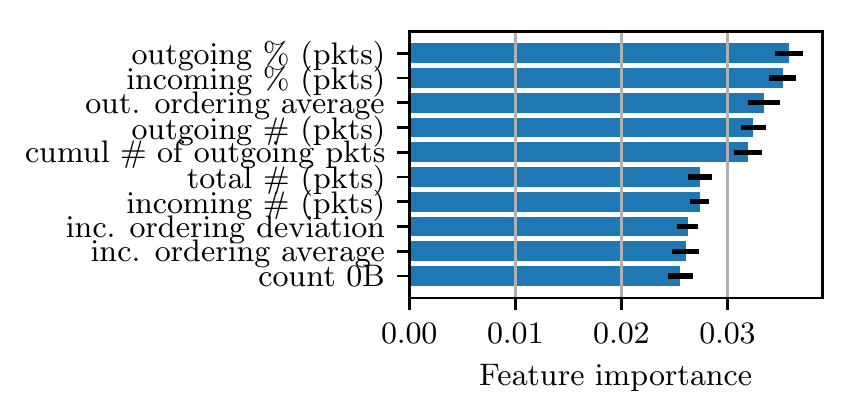}
	\vspace{-0.2cm}
	\caption{Feature importance when hiding global features (last row of~\autoref{table:quic-defenses-full-view}).}
	\label{fig:sizes-volumes-hidden}
	\vspace{-0.4cm}
\end{figure}

\parait{Hiding global features.}
To hide the total transmitted volume, we increase the size of all packets such that the total transmitted size is padded to the next megabyte.
This yields another small drop in performance.
The attacker simply starts using packets orderings as main feature (\autoref{fig:sizes-volumes-hidden}), which are almost as informative as sizes.

\parait{Injecting dummies.}
We then include dummy packets (padded to the maximum size) to hide individual packet orderings.
Since we cannot use existing defenses based on adversarial perturbation to find the optimal placement of dummies (see \autoref{sec:background}), we inject dummy packets based on FRONT~\cite{gong2020zero}, which ~\cite{smith2022qcsd} showed can be implemented as a client-side-only QUIC defense.
FRONT has four parameters that can be adjusted: $N_s$ and $N_c$, which are the maximum number of dummies that can be sent by the server and client respectively, and $W_{min}$ and $W_{max}$, which indicate the time window within which the dummies must be sent.

We inject dummies using the FRONT parameters suggested by Smith~\etal ~\cite{smith2022qcsd}: $N_s = N_c = 1300$, $W_{min} = 0.2s$, $W_{max} = 3s$.
We calculate the bandwidth overhead using the same definitions as in ~\cite{smith2022qcsd}, \ie the increase relative to the transmitted data. 
We find that FRONT achieves a significant reduction only at a sharp increase in cost: to obtain $\approx 70\%$ F1-score drop for the adversary requires an overhead of 2.31. 
This is in addition to the already large overhead in terms of padding used to hide local and global features (mean cost of $612$ kB per trace, with a large standard deviation: $440$ kB). 
We also experiment with different FRONT parameters (as shown in Table~\ref{table:front}), and find that reducing the overhead comes at the cost of defense effectiveness.
(We note that we cannot have a direct comparison with Smith~\etal~\cite{smith2022qcsd} since their evaluation is in an open-world VPN scenario.) 

\begin{table}
	\caption{F1-score and bandwidth overhead when injecting dummies using FRONT. We vary the FRONT parameters $N_s$, $N_c$, and $W_{max}$.}
	\centering
	\begin{tabular}{l | l | rrr}
		\toprule
		\textbf{F1-Score/Overhead} & & \multicolumn{3}{c}{$N_c = N_s$}\\
		\midrule
		&  & 325 & 650 & 1300 \\ 
		\midrule
		\multirow{3}{2em}{$W_{max}$}  & 0.5 & 64.9/0.55 & 60.6/1.15 & 58.3/2.24 \\
		& 1 &  60.4/0.55 & 53.8/1.12 &  48.1/2.26 \\
		& 2 & 53.3/0.56 & 43.3/1.12 & 34.9/2.22 \\
		& 3 & 51.4/0.56 & 41.0/1.12 & 31.4/2.31 \\
		\bottomrule
	\end{tabular}
	\label{table:front}
\end{table}

\begin{table}
	\caption{\quicM dataset: Mean classifier performance on defended traces.}
	\centering
	\begin{tabular}{lrrrr}
		\toprule
		\textbf{Variant} & \textbf{F1-Score} & \textbf{Std.\ dev.} \\
		\midrule
		undefended & 95.8 & 0.4 \\
		hiding individual sizes & 93.9 & 0.4 \\
		hiding all timings & 95.5 & 0.3 \\
		+ hiding total transmitted sizes & 92.2 & 0.5 \\
		\bottomrule
	\end{tabular}
	\label{table:quic-defenses-full-view}
\end{table}

\label{sec:limadv}
\subsection{Constrained Adversary}
As described in \autoref{sec:model}, there are on-path adversaries who might not observe all traffic from a particular client, nor all traffic to a particular server of interest;
or that may not have the capability to process this traffic or to transmit it to a location suitable for analysis.
We now study the performance of these adversaries, and whether \texttt{PADDING}-based network-layer defenses could protect against them. 
We run our experiments on the \quicM dataset.

\subsubsection{Limited traffic visibility}
\label{sec:adv-partial-view}

To understand the impact of limited visibility on the adversary's performance, we simulate an AS adversary with partial view of the client's traffic.
We determine which parts of the traffic an AS would see using HAR captures to identify resources requested during page loads.
Then, as in prior work~\citep{juen2015defending}, we use traceroutes to record the ASes in the path taken by the resource requests. 
Concretely, we set \texttt{traIXroute}~\citep{nomikos2016traixroute} to use \texttt{scamper} (configured with the Paris traceroute technique).
As Juen \etal~\citep{juen2015defending}, we discard route hops that do not have IP or AS information (asterisks in the traceroute).
To avoid inaccuracy in our analysis, we do not attempt to fill these gaps in routes via stitching. 
Thus, our results provide a lower bound on the amount of traffic that an AS adversary sees.

We then simulate the partial view of the adversary by filtering out TLS/QUIC connections that do not correspond to the resources visible to the adversary.
This filtering is based on the destination IP address, and the SNI when it exists.
We observe a total of \numtraceroutes routes in the \quicM dataset.
These results are from traceroutes collected from the same location as the \quicM dataset (in France).
We ran traceroutes from other vantage points (Germany, UK, and Singapore), and observed the same trends.
We report on these additional experiments in Appendix~\ref{app:vantage}.

\begin{figure}
	\centering
	\small
	\includegraphics[width=\linewidth]{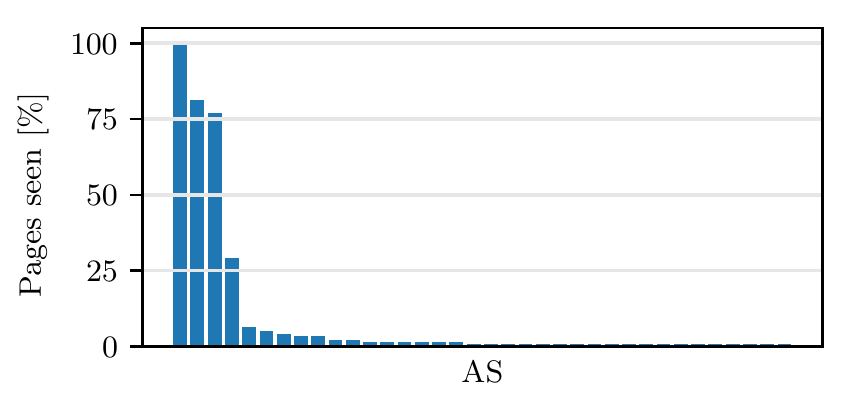}
	\vspace{-0.2cm}
	\caption{Distribution of web pages seen by each AS.\@ Only three ASes (client's AS, Google, Cloudflare) can observe traffic from all the pages. Most ASes observe less than 10\% of pages.}
	\label{fig:asn_seen_numsites}
	\vspace{-0.4cm}
\end{figure}

\begin{figure}
	\centering
	\small
	\includegraphics[width=\linewidth]{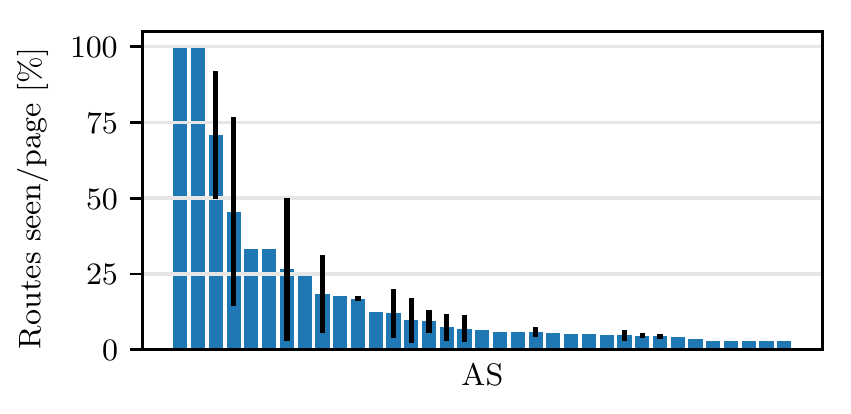}
	\vspace{-0.2cm}
	\caption{Distribution of routes per web page seen by each AS.\@ Only three ASes (client's AS, OVHcloud, Google) observe more than 50\% of the traffic per site.}
	\vspace{-0.4cm}
	\label{fig:asn_seen_persite}
\end{figure}

\autoref{fig:asn_seen_numsites} shows how many webpages are \emph{seen} by each AS. 
We consider the web page is seen if the AS sees any traffic associated with this web page visits (including its subresources).
There are three ASes that observe traffic from more than 80\% of webpages: one from Google, one from Cloudflare, and the AS where our client is located. 
The majority of the ASes, however, see only a small proportion of the sites (less than 10\%).
If these ASes were to be the adversary, they would not be able to fingerprint traffic from most websites hosted by the IP our attacks target.\@

We show the classifier performance for some ASes in \autoref{table:AS-view}. 
The few ASes that have a substantial view of the client connections, \eg Google or Cloudflare, obtain high F1-score.
Most other ASes observe traffic from very few pages, e.g., LEVEL3, Facebook, or VNPT-AS-VN observe less than 5\% of the pages in the dataset, meaning that they cannot fingerprint the remaining 95\%. 
For the pages that they observe, they obtain high F1-scores.
VNPT-AS-VN observes traffic for just one page and, thus, always identifies it. 

We conclude that, in order to successfully fingerprint, an AS adversary needs to observe a large proportion of the traffic, either by being the client's AS or by providing sub-resources on websites.
For every web page an AS sees, we study what portion of this page they can observe (see \autoref{fig:asn_seen_persite}). 
Our source AS, naturally, sees 100\% of the traffic of all pages in our dataset. 
Another AS, 4367 (belonging to OVHcloud), sees 100\% of page traffic for a very limited number of pages.
The Google AS is the second highest, seeing $\approx$ 70\% of the routes for $80\%$ of the pages in the dataset. 
All other ASes see less than 50\% of the routes associated with the pages for which they can observe traffic and therefore are not much of a threat.

\begin{table}
	\caption{Mean classifier performance on different AS views.}
	\centering
	\begin{tabular}{lrrrr}
		\toprule
		\textbf{AS} & \textbf{Name} & \textbf{\# Pages} & \textbf{F1-Score} \\
		\midrule
		\texttt{15169} & Google, LLC & 118 & 89.5 \\
		\texttt{13335} & Cloudflare, Inc & 115 & 92.9 \\
		\texttt{3356} & LEVEL3 & 7 & 81.7 \\
		\texttt{32934} & Facebook, Inc & 5 & 92.3 \\
		\texttt{45899} & VNPT-AS-VN & 1 & 100.0 \\
		\bottomrule
	\end{tabular}
	\label{table:AS-view}
\end{table}

Given these results, we expect the effectiveness of network-layer \texttt{PADDING}-based defenses in this scenario to be similar to the unconstrained adversary. 
If the AS has high visibility on the traces, they are essentially unconstrained and defenses cannot significantly reduce the performance.
If the AS observes little traffic, their performance will be already low and the gain provided by defenses can only be marginal, while still imposing high bandwidth overhead.

\subsubsection{Limited processing power}
\label{sec:adv-realistic-computation}

To perform website fingerprinting, adversaries must have storage and computation capabilities, which middleboxes typically do not have.
In fact, typical network monitoring solutions only record aggregate statistics over sampled traffic~\citep{silva2017inside}. 
Common tools for network sampling include \emph{NetFlow} and \emph{sFlow}~\citep{murdoch2007sampled}.
More efficient techniques have been proposed in academic papers (\eg sketching~\citep{krishnamurthy2003sketch, liu2016one} or \emph{skampling}~\citep{tune2014ofss}), but to the best of our knowledge, they are not widely used.

We focus on NetFlow, which is widely deployed on the Internet.
We simulate \textit{Sampled NetFlow}, a variation of NetFlow used in high-speed links where packets are first sampled in a deterministic fashion (1 out of every N packets) and flow statistics are computed on the sampled packets.
We down-sample the PCAPs uniformly to the desired sampling rate, and then we create NetFlow summaries from the PCAPs using \texttt{nfpcapd} and \texttt{nfdump}. 
We experiment with various sampling rates: $100\%, 10\%, 1\%$ or $0.1\%$ (common sampling rates in the wild range from 50\% to 0.1\%~\citep{carela2011analysis}).
Then, we adapt the features used by our classifier to NetFlow summaries.
These summaries record the number of packets but not their sizes, timings, or directions.
For classification we consider a flow as a single packet whose size is the sum of all packets in a flow, and inter-packet timings become inter-\emph{flow} timings.
We acknowledge there could be better features and that our evaluation only provides a lower-bound on the attacker performance.

\begin{table}
  \centering
  \caption{Mean classifier performance and median storage cost per sample for Sampled NetFlow, \quicM.}
  \vspace{-0.2cm}
  \begin{tabular}[c]{  c  c  c }
    \begin{tabular}{lrrrr}
    	\toprule
    	\textbf{Sampling} & \textbf{F1-Score} & \textbf{Size [kB]} \\
    	\midrule   
    	Full traces & 95.8 & 312.4 \\
    	NetFlow 100\% & 90.5 & 25.9 \\
    	NetFlow 10\% & 66.4 & 3.0 \\
    	NetFlow 1\% & 41.7 & 0.9 \\
    	NetFlow 0.1\% & 16.8 & 0.4\\
    	\bottomrule
    \end{tabular}
  \end{tabular}
  \label{table:netflows-perf-storage}
  \vspace{-0.4cm}
\end{table}

\begin{figure}
\centering
\includegraphics[width=\linewidth]{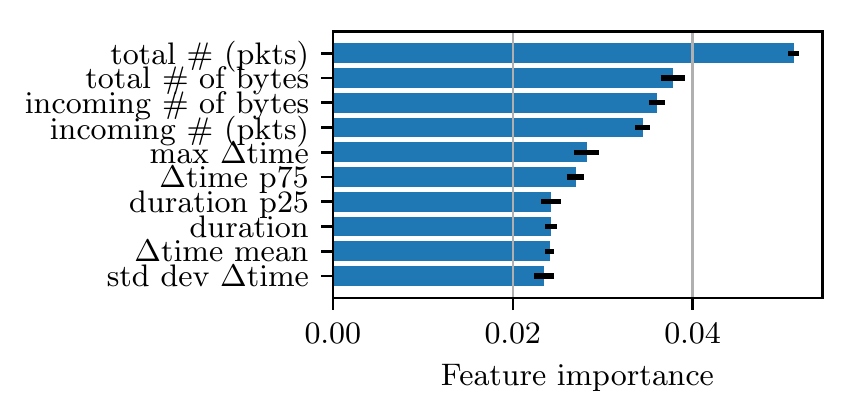}
\vspace{-0.6cm}
\caption{Feature importance for classifying NetFlow with $1\%$ packet sampling rate.}
\vspace{-0.4cm}
\label{fig:netflows-fi}
\end{figure}

We show the adversary's success on the NetFlow summaries in \autoref{table:netflows-perf-storage}.
Moving from full packet data to flow summaries leads to a significant reduction in the adversary's performance: 
$29$ percentage points lost when $10\%$ of packets are sampled.
Yet, the F1-score for any sampling rate is much higher than random guessing (F1-score=$0.6\%$).

\begin{table}
  \centering
  \vspace{-0.2cm}
  \caption{Mean attacker performance on defended NetFlow, \quicM.}
 \vspace{-0.2cm}
  \begin{tabular}[c]{ l  c  c }
   \toprule
        \textbf{Sampling} & \textbf{F1-Score} \\
    \midrule
    NetFlow 100\% & 53.1 \\
    NetFlow 10\% & 33.1 \\
    NetFlow 1\% & 21.6 \\
    NetFlow 0.1\% & 8.6 \\
    \bottomrule
  \end{tabular}
 \vspace{-0.4cm}
 	\label{table:netflows-defense}
\end{table}

\parait{Defending NetFlow traces}
Regardless of the sampling rate, the most important features when using NetFlow are the total number of bytes and packets (\autoref{fig:netflows-fi}).
We explore a defense that hides both per-flow metrics and overall statistics about the number of bytes and packets exchanged.
For full traces, we hide global statistics by padding the total transmitted bytes to $22$ MB and the number of packets to $25$K (the maximum transmitted size and number of packets we observe in our dataset).
This padding is added uniformly to all the flows of one sample.
For sampled traces, we reduce the padding with the sampling rate.

The defense reduces the attacker performance, introducing prohibitive cost ( $\approx39$ MB per trace in median, see~\autoref{table:netflows-defense}).
We also observe that most of the gain in privacy compared to the standard setting ($95.8\%$) comes from the sampling rather than the padding (e.g., for Netflow 1\%, $-54.1$\% via sampling vs. $-20.1$ \% via padding).

\subsubsection{Inexpensive fingerprinting due to resource centralization}\label{sec:google-attack}

We now show how the common use of Google resources by web developers can be used by constrained adversaries to bypass these limitations.

From the (incomplete) AS information on the network level, we found that at least $118$ out of the $150$ websites in our dataset were seen by Google, i.e., the traffic traverses Google's AS.
From the HAR captures, we observe that most websites request resources from a Google-owned domain ($125$ websites in \quicM ($83\%$ of the dataset)).
We confirm this result in Section~\ref{sec:app-structure-analysis}.
While studying the traces, we also observe that the order in which these resources are loaded is website-particular, i.e., even when two sites load the same set of resources, these resources are loaded at different times --with respect to the time of query of the home page-- and in different orders. 
Such website-dependent behaviour results in a fingerprint.
This fingerprint, caused by the centralization of web resources at Google, can be used by an adversary to perform traffic analysis at a fraction of the typical cost: instead of recording all traffic, an adversary can use the timings of \texttt{ClientHello}'s to Google IPs. 
%

To validate this, we study the performance of an adversary that only records the traffic towards Google services.
We filter the network traces for which the destination IP (or the SNI) belongs to Google. 
If this field is not present in the packet, we perform a reverse-mapping with the destination IP to confirm the destination.
To list domains belonging to Google, we check the ownership of the requested URLs in our HAR capture using Tracker Radar~\citep{tracker-radar}.
We use the following Google-owned domains: \texttt{google.com}, \texttt{gstatic.com}, \texttt{youtube.com}, \texttt{doubleclick.com}, \texttt{ggpht.com}.
We extract the sending times of the packets containing a \texttt{ClientHello} to these Google IPs and domains.
For \quicM, this represents $7.6$ floating-point values on average per loading of a website, with a maximum of $27$ values.
The size of the fingerprint is between $61$B and $216$B per loading of a website; in contrast, the mean .pcap size is $112$ kB for the traffic towards Google, and $312$ kB for all traffic. 

\begin{table}
  \centering
  \caption{Mean classifier performance and median storage required per sample on traces filtered by connections to Google services, \quicM. The last two rows use 125 samples.}
  \begin{tabular}{lrrrr}
  	\toprule
  	\textbf{Variant} & \textbf{F1-score} (Std dev) & \textbf{Size [kB]} \\
  	\midrule
  	Baseline (Full Traffic) & 95.8 (0.4) & 312.4 \\
  	Full traffic to Google & 78.4 (0.7) & 112.1 \\
  	\texttt{ClientHello}'s to Google & 66.1 (0.4) & 0.1 \\
  	\bottomrule
  \end{tabular}
	\label{table:google-dest-filter}
	\vspace{-0.4cm}
\end{table}

\begin{figure}
	\centering
	\includegraphics[width=\linewidth]{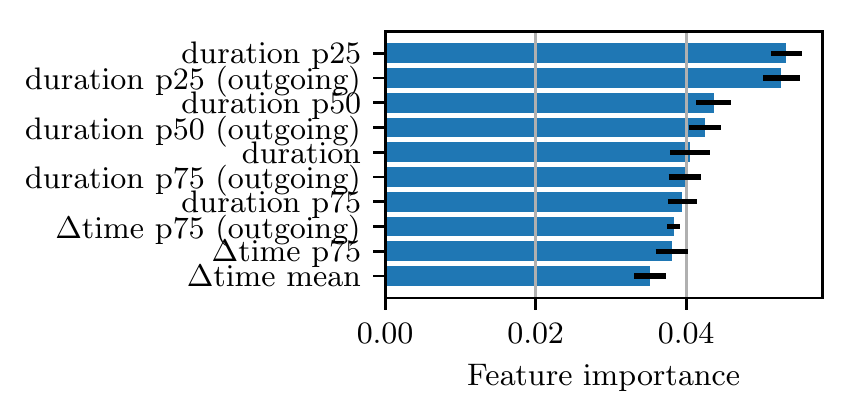}
	\vspace{-0.6cm}
	\caption{Feature importance for classifying websites based on the timings of their requests to Google services.}
	\vspace{-0.4cm}
	\label{fig:google-dest-filter-fi}
\end{figure}

We show in \autoref{table:google-dest-filter} that even when \emph{using only the timing of requests to Google}, the adversary achieves a F1-score of $77.9\%$ for the 125 websites that use some Google resource.
To obtain this result, the adversary needs \emph{only $\approx 61$ B per connection}, a saving of four to five orders of magnitude compared to recording full network traces.
The feature analysis confirms that the timings between sub-resources is what helps the attacker in this case (\autoref{fig:google-dest-filter-fi}).

\subsection{Take-aways}

Our experiments confirm that website fingerprinting is a threat for QUIC traffic~\cite{smith2021website}, both when users browse landing pages and when they visit subpages.
While factors such as time and user client can influence the adversary's performance, these can be overcome by retraining the classifier with newer data, and by using classifiers tailored to specific client setups.
We show that the threat persists even when the adversary has limited visibility or storage capabilities.
Interestingly, the fact that resources are centralized in a few CDNs can enable other entities to perform fingerprinting at a fraction of the usual cost: \eg an ISP can use the Google-filter to save bandwidth by $3$ to $4$ orders of magnitude (compared to running the attack on all traffic).

We find that network-layer \texttt{PADDING}-based defenses are not very effective at thwarting website fingerprinting attacks. 
Despite their high cost, they can only reduce an unconstrained adversary's success by, at most, $3.5$ percentage points without dummies. 
Dummy injection increases protection, at a prohibitive cost: a 2.31 overhead to achieve a 70\% performance reduction.
Network-layer \texttt{PADDING}-based defenses can help against bandwidth-constrained adversaries, but the bulk of the privacy gain  stems from the adversary's need to sample rather than from the padding. 

The failure of network-layer \texttt{PADDING}-based defenses is because the anonymity sets behind IPs are usually small, and the classifier is able to pick even small differences between global statistics of the traces such as total transmitted size or the total number of packets that will be sent. The defenses we study cannot hide these differences because, \emph{at the network layer, they lack information on these global statistics}.

\section{Designing application-aware \texttt{PADDING}-based defenses}
\label{sec:applayer}

The results in the previous section confirm that the findings of Dyer \etal~\citep{dyer2012peek} for HTTP over encrypted tunnels hold for QUIC traffic: network-layer defenses have limited ability to hide global traffic patterns. 
In this section, we explore whether \textit{application-aware defenses} can be more effective against website fingerprinting.

\para{Application-aware defenses.} 
We consider any defense that requires knowledge of a page's resources to be an application-aware defense. 
This can be knowledge of the number of resources, their size, and their order; or knowledge about global information such as the total (incoming or outgoing) size.
The only way to obtain this information is to extract it from the application layer, as Smith \etal~\cite{smith2022qcsd} do to implement FRONT~\cite{gong2020zero} and Tamaraw~\cite{cai2014tamaraw} in practice.
However, while their work demonstrates that QUIC being implemented in the user space makes it possible to implement existing network-layer defenses using a user-space library, 
they did not explore whether application-layer information could be used to tailor how dummies are injected into a trace. 
In contrast to Smith \etal~\cite{smith2022qcsd}, we aim to understand whether using application-layer information defenses fare better than their uninformed network-layer counterparts, and what would it take to deploy such defenses.

In this section, we analyze the structure of websites to gain understanding about what information can be extracted and used to inform defenses.
Second, we study different ways in which this information can be used to build defenses, and evaluate how these defenses perform.

\subsection{Understanding web page composition}
\label{sec:app-structure-analysis}
We study different dimensions related to the structure and composition of websites that are relevant for configuring \texttt{PADDING}-based defenses.
%
We collect page structures by crawling the pages in \quicM with OpenWPM (v0.17.0)~\citep{openwpm} using Firefox five consecutive times.
OpenWPM logs the HTTP requests that occur during the page load.
Unlike HAR captures, OpenWPM also records the originator of a request.

\para{Resource dynamism.} We first study how dynamic the pages in our dataset are.
Dynamism influences how easy it is to protect a page.
The less dynamic pages are, the easier it is to protect them as one can select defense parameters tailored to the static resources.
If pages vary overnight, defenses can only be configured to fit the average case.

\begin{figure*}[!htpb]
	\centering
	\begin{subfigure}{0.32\linewidth}
	\includegraphics[width=\textwidth]{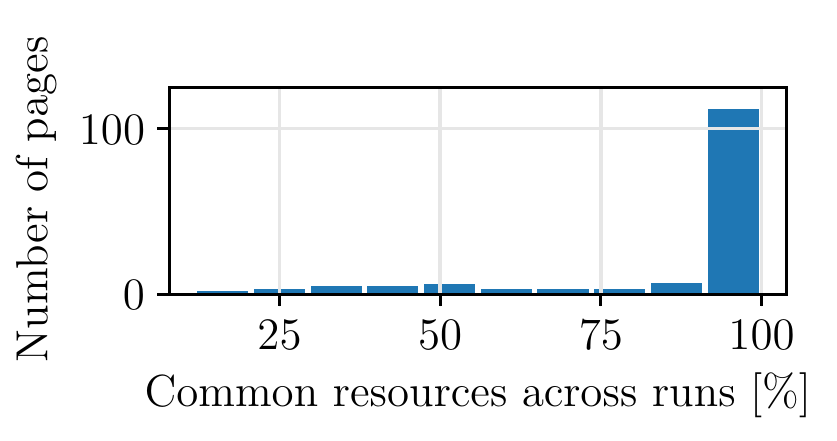}
	\vspace{-2em}
	\caption{\label{fig:rstatic}}
	\end{subfigure}
	\hfill
	\begin{subfigure}{0.32\linewidth}
	\includegraphics[width=\textwidth]{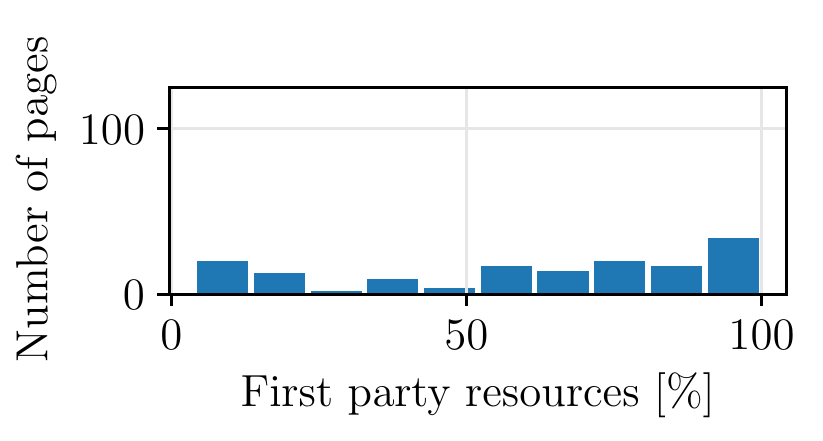}
	\vspace{-2em}
	\caption{\label{fig:firstparty}}
	\end{subfigure}
	\hfill
	\begin{subfigure}{0.32\linewidth}
	\includegraphics[width=\textwidth]{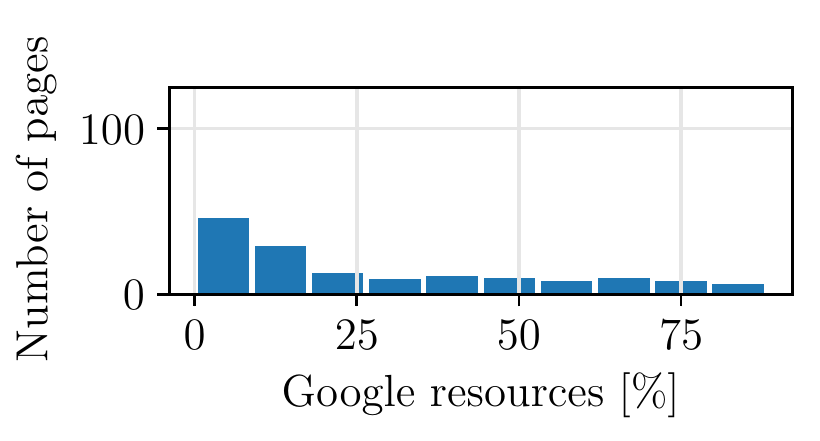}
	\vspace{-2em}
	\caption{\label{fig:google-tp}}
	\end{subfigure}
	\vspace{-0.6em}
	\caption{Resource dynamism and ownership for pages in the \quicM dataset. (a) Distribution of the proportion of resources that remain static across 5 runs.
The majority of resources remain constant across runs, indicating low page dynamism. (\subref{fig:firstparty}) Proportion of first party resources. 18\% of the pages have less than 20\% of first-party resources. (\subref{fig:google-tp}) Proportion of Google third party resources. 24\% of the pages have more than half their resources served by Google.}
	\vspace{-0.6em}
	\label{fig:resources}
\end{figure*}

Out of the 150 websites in \quicM, 149 were successfully visited across all crawls. 
For these pages, we calculate the proportion of resources that remain static across the runs.
Sometimes, even if the resources fetched are the same, the URL parameters may vary.
We strip the URL parameters, and plot the distribution of static resources in \autoref{fig:rstatic}. 
The mean proportion of static resources is 88.25\% (Std: 22.46\%) and the median is 100\%.
This indicates that pages in our dataset mostly contain static content.
We note, however, that our measurements are taken over a short period of time and dynamism could become more prevalent when web pages are observed over a longer time period.  

\newcommand{\Rstatic}{R_{\text{s}}}

\para{Resource ownership.}
Resource owners are the ones that can provide information about resources and modify them.
Understanding ownership is important to get an idea of how much coordination among owners is required to protect a page.
We study whether resources belong to the \textit{first} party (same eTLD+1 as the page) or to a \textit{third} party (different eTLD+1 from the page).
For example, on the page \texttt{www.example.com},  \texttt{img.example.com} would be first-party and  \texttt{external.com} would be third-party.
Using domains as a proxy for ownership is not perfectly accurate: content for \texttt{facebook.com} can be served from \texttt{fbcdn.net}.
While both domains come under Facebook's control, the latter would be identified as a third party. 
Unfortunately, we need to rely on domains as existing services that provide entity-domain mappings~\citep{tracker-radar} do not have relevant ownership information for almost half of our dataset. 

\autoref{fig:firstparty} shows the proportion of first-party resources for our dataset. 
The majority of the pages have a large proportion of first-party resources (Mean 61.18\%, Std 31.61\%, Median 66.67\%), though some pages have as few as 3.92\%. 
On average, there are 5.95 third party domains per page (Std: 7.64, Median: 3), with the number of domains going up to 44 for one of the pages.
Visual inspection of the third-party domains shows a large number of domains commonly associated with Google.
We map the domains to their owning entities~\citep{tracker-radar} to measure Google's prevalence (\citep{tracker-radar} contains mappings for Google's domains).
Figure~\ref{fig:google-tp} shows the proportion of Google resources per page.
24\% of the pages have more than half their resources served by Google.

\subsection{Application-aware defense strategies}
\label{sec:app-layer-defenses-strat}

\subsubsection{Party-based resource protection}
The adversary can perform the attack by filtering resources from different origins (\eg only on Google resources as in Sec~\ref{sec:google-attack}).
We assess whether third parties need to be involved or if only protecting first-party resources suffices.

We build traces with only first- or third-party resources, and protect those using padding.
First-party protection represents a scenario where web pages protect their content using some defense, but third-parties do not cooperate.
Third-party protection represents a scenario where third parties such as CDNs, which host a large number of resources, decide to implement a defense, but smaller first-parties do not.
We assume the adversary attacks the remaining undefended traffic.
As shown in \autoref{table:har-quic150-attack-by-parties},
the adversary can achieve a high performance by just analyzing partial, undefended traces, regardless of the origin.
Thus, for any defense to be effective, all parties serving content for a page must coordinate and actively participate in the protection of resources.
In particular, due to its prevalence, \emph{Google must collaborate for any \texttt{PADDING}-based defenses to be efficient}.

\begin{table}
	\caption{Mean classifier performance on traces filtered by 1\textsuperscript{st} / 3\textsuperscript{rd} party and Google CDN, \quicM.}
	\centering
	\vspace{-0.2cm}
	\begin{tabular}{lrrrr}
		\toprule
		\textbf{Variant} & \textbf{F1-Score} & \textbf{Std. dev} \\
		\midrule
		All traffic & 0.937 & 0.008 \\
		Only traffic to/from 1\textsuperscript{st} parties & 0.955 & 0.004 \\
		Only traffic to/from 3\textsuperscript{rd} parties & 0.915 & 0.005 \\
		Only traffic to/from Google CDN & 0.914 & 0.006 \\
		\bottomrule
	\end{tabular}
	\vspace{-0.4cm}
	\label{table:har-quic150-attack-by-parties}
\end{table}

\newcommand{\padres}{\texttt{pad}$_{\text{resources}}$}
\newcommand{\padtot}{\texttt{pad}$_{\text{total size}}$}

\subsubsection{Evaluating application-aware defense strategies}
\label{sec:app-layer-defenses-eval}

Application-aware defenses can be implemented both at the application layer or at the network layer (if information is passed to the middleware implementing padding).
We directly evaluate perturbed application-layer traces as this gives an upper bound on the performance of a network-layer adversary with respect to a set of features~\citep{hentgen2019measuring}. 
The reason is that network-layer features are effectively a noisy version of application layer resources~\citep{hentgen2019measuring} (e.g., the number or total size of incoming QUIC packets are a noisy version of the actual size of the downloaded resources, and the total duration of the connection is a noisy version of the total amount of bytes downloaded).
We study three scenarios: undefended traces with all features, undefended traces without timings, and defended traces without timings.
The latter is a good estimate of the attacker performance (even with timings), as our baseline analysis shows (see Section~\ref{sec:app-def-eval}).

\parait{Metrics.}
We use two metrics to evaluate the defenses' success: the performance of the classifier and the overhead imposed in terms of kilobytes of data added per subrequest.

\parait{Dataset and features.}
For the undefended baseline, we use the HAR captures of \quicM.
We derive the k-Fingerprinting features of these captures (which output a list of tuples $[\text{t}_\text{request}, \text{size}_\text{request}, \text{t}_\text{response}, \text{size}_\text{response}]$).

In practice, our padding and dummy-injection defenses would affect the timing of packets.
However, without deploying the defenses we cannot predict these changes would reflect on our traffic captures.
Deployment is not a possibility, as even if we would copy all websites of \quicM on a server we control, we would not be able to simulate actions from third parties.
Fortunately, timings are less stable (and hence, less useful) than sizes, and therefore, they are not among the most important features (~\autoref{fig:har-quic150-fi} in Appendix~\ref{app:feature_importance}).
In fact, when attacking full HARs, the adversary obtains very good performance ($93\%$ F1-score) both with and without timing information.
The most important features are size-related, being \texttt{bytes\_outgoing} the most important feature by a slight margin over \texttt{bytes\_incoming} (\texttt{bytes\_total} is the sum of the two).
This is corroborates the findings by Hentgen, who shows that even without timings evaluating at the application layer yields an upper bound over the network layer~\citep{hentgen2019measuring}.
In the remaining experiments, we discard timings.

\subsubsection{Defense strategies.}
\label{sec:app-def-eval}

We now evaluate possible strategies that use application-layer information to decide how to configure \texttt{PADDING} frames.

\parait{Protecting local features with padding.}
We design a padding function \padres~to hide individual queries and resources sizes.
Such a defense must be implemented both on the client and the server.
To configure the function, we use (1) the distribution of resource sizes in the set of websites and (2) a parameter $N$, which defines how many different sizes the defense allows for.
The padding function splits the resources sizes into $N$ groups of equal density.
For instance, if $N=1$, all resources are padded to the max resource size in \quicM; and if $N=2$, half of the resources are padded to the median size, half to the max size.
Choosing a small $N$ increases privacy: more resources will be padded to the same size and be indistinguishable; but also increases bandwidth usage.

\begin{figure}
	\centering
	\includegraphics[width=\linewidth]{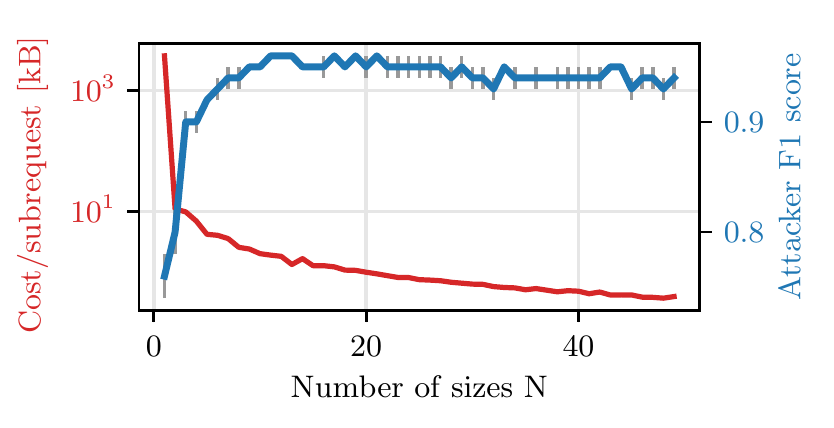}
	\vspace{-2em}
	\caption{Number of sizes, $N$, in \padres versus attacker performance and median bandwidth cost per subrequest.}
	\vspace{-1.6em}
	\label{fig:har-quic150-pad-steps-vs-acc-cost}
\end{figure}

We run this defense varying $N$, and plot the median cost and the attacker F1-score in~\autoref{fig:har-quic150-pad-steps-vs-acc-cost}.
Only large amounts of padding (small $N$) have an effect on the attacker accuracy.
Padding with large sizes has little effect.
For instance, $N=3$, which results on packets of $5.58$ kB, $21$ kB, $3.6$ MB, decreases the accuracy adversary by $6\%$ and incurs a median overhead of $9$ kB per resource.
This ineffectiveness stems from the fact that the adversary still has access to the number of requests and overall volume (see \autoref{fig:har-quic150-pad-3steps-fi} in Appendix~\ref{app:feature_importance}), which are sufficient for the attack.
The traces' total size are too different to be efficiently hidden through the padding of individual resources.

\parait{Protecting global features with padding.}
Padding only individual packet sizes cannot protect the overall transmitted volume.
We design a padding function \padtot~to pad the total incoming and outgoing sizes.
To evaluate the best case defense, we assume the ideal scenario in which the padding effort is split evenly across all the parties queried on one web page.
This way, the adversary does not gain an advantage by filtering the traces from one party in particular.
This strategy assumes the existence of a mechanism by which clients can ask third parties for a particular amount of padding per resource, e.g. using the method by Smith \etal~\cite{smith2022qcsd}.

The defense has one parameter, $N$, which defines the total incoming and outgoing traffic sizes that are allowed.
We first compute the maximum total incoming and outgoing traffic in our target dataset, \quicM. 
The maximum total size of queries per website is $102$ kB and the median is $14.4$ kB; and the maximum total size of all downloaded resources is $8.19$ MB, with median $750$ kB. 
To apply the defense, we split the total incoming/outgoing sizes into $N$ groups of traces with equal density.
For instance, when $N=1$, there is only one group of maximum size: all websites' outgoing traffic would be padded to $102$ kB, and the incoming traffic to $8.19$ MB. 
For $N=2$, the groups would correspond to the median and to the maximum total incoming and outgoing traffic.
For $N=3$, the groups would correspond to tertiles of the distribution, and so forth.
We allocate every website to the group that is closest to its original total incoming and outgoing size, and we spread the padding evenly across all queries and resources of that website.

\begin{figure}
	\centering
	\includegraphics[width=\linewidth]{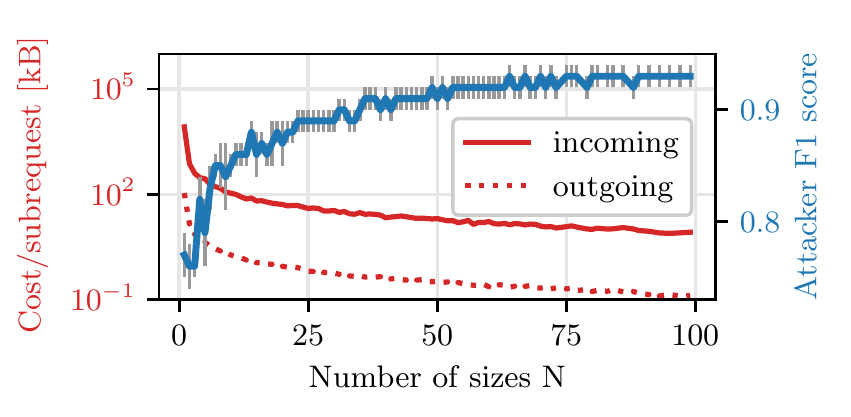}
	\vspace{-2em}
	\caption{Number of sizes $N$ in \padtot versus attacker performance and median bandwidth cost per subrequest.}
	\vspace{-1em}
	\label{fig:har-quic150-pad-total-cost-accuracy}
\end{figure}

We run this defense varying $N$, and plot the median cost and the attacker F1-score in~\autoref{fig:har-quic150-pad-total-cost-accuracy}.
As in the network layer, padding the total size does not mitigate the attack.
For instance, to drop the adversary's accuracy by $10$ percentage points, the defense incurs a median cost of $5.7$ kB per request (outgoing traffic) and $300$ kB per resource (incoming traffic).
In the best case, it reduces the attacker's accuracy by $16$ percentage point, with a median cost of $109$ kB per request and $8.16$ MB per resource.

\parait{Protecting global features with dummies.}
Injecting dummy traffic can also hide the total size~\citep{juarez2016toward}.
Unlike in Tor, care must be taken that dummies' sizes do not enable the adversary to filter them.
Considering ideal circumstances in which the adversary cannot filter dummies, we assume dummies are sampled from a dataset that contains the most popular queries and resources across all pages to be defended. 

In our experiments, we select popular resources from Google (fonts, analytics, static assets). 
When a web page is loaded, we choose a number of resources to inject ($M$).
These resources can themselves trigger additional queries.
We flip a coin and with probability $p$, we inject a dummy resource.
We inject these resources at random times over the duration of the connection, such that the adversary cannot use timing to identify and filter out dummies. 

\begin{figure}
	\centering
	\includegraphics[width=\linewidth]{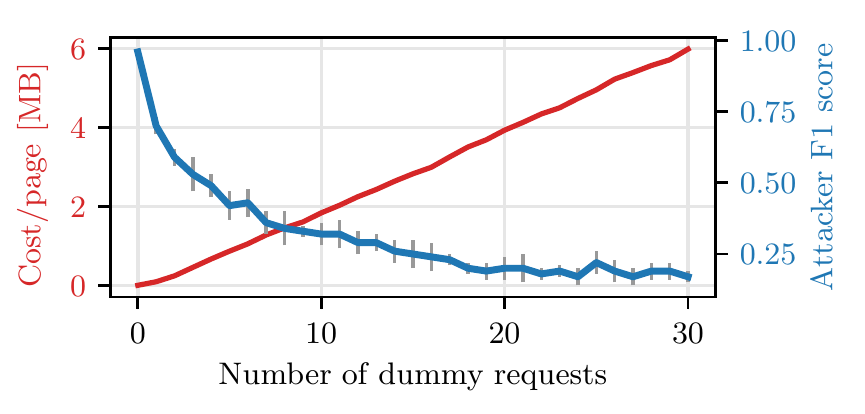}
	\vspace{-2em}
	\caption{Attacker performance and cost when varying the number of dummies.}
	\vspace{-1em}
	\label{fig:har-quic150-dummies}
\end{figure}

We plot in~\autoref{fig:har-quic150-dummies} the attacker F1-score for a varying number of dummies.
This defense is more effective than the previous ones. 
For instance, with $(p=0.5, M=10)$, which injects on average $5$ dummy requests, the attacker's F1-score decreases from $93\%$ to $54\%$, at a median cost of $137$ kB per loading of a web page.
In general increasing $p$ has better impact on the attacker's performance than increasing the quantity of dummies ($M$): on average, the two parametrizations $(p=0.5, M=10)$ and $(p=1, M=5)$ inject $5$ sequence of queries to a CDN; but the former reduces the attacker's F1-score to $0.54$, and the latter to $0.43$.

\subsection{Take-aways.}
Padding resources at the application layer incurs large overhead and does not necessarily eliminate website fingerprinting attacks. Dummy injection, on the other hand, shows better trade-offs.
On the downside, contrary to pure network-layer defenses, when information is protected at the application-layer, there is a need for coordination between first and third parties. 
Moreover, application-aware defenses need a-priori knowledge of the resources sizes and ordering.
This may be hard to obtain when browser \& website optimizations (\eg client caching or pipelining of resources) are in place. 
Thus, the deployment of these defenses may require the collaboration of web pages and web browsers developers besides the coordination of parties that serve web content.

\section{Discussion and Recommendations}
\label{sec:conclusion}
We carried out a comprehensive study of the ecosystem in which defenses are to be deployed, considering the majority user that does not use Tor/VPNs.
We provided evidence of fundamental incompatibilities between today's Web practices and the deployment of effective defenses.

\para{Network-layer challenges.}
First, we confirm that network-layer defenses are not effective against website fingerprinting~\citep{dyer2012peek}, and show that this also applies when the transport protocol changes from TCP to QUIC.\@
The main problem stems notably from the differences in the total sizes of websites which result in identifying features~\cite{hayes2016kfingerprinting,OverdorfJAGD17}.
Hiding the total size is hard at the network layer, where the size of objects is not known in advance.
Thus, without coordination with the application layer, using the QUIC's \texttt{PADDING} frame is unlikely to result on an effective defense against traffic-analysis attacks.
Effective mitigations require application involvement, either in the application code or as part of the browser's functionalities~\cite{davidson2022tango}.

\para{App-layer challenges \& Next steps.}
While defenses at the application layer can obtain better protection at a smaller cost than network-layer defenses, our investigation shows that the current Web development practices hinder the effective deployment of any defense.
Websites use resources hosted on different servers, and defenses must cover \emph{all} resources to be effective (see \autoref{sec:app-layer-defenses-strat}).
The widespread use of third party resources means that achieving full coverage requires coordination among many different entities, which seems unlikely to happen organically.

To improve the situation without the need for coordination between first and third parties, Web-oriented standard bodies (\eg W3C) and browser vendors could develop mechanisms to standardize how third-party resources are requested and served.
For instance, defining standard sizes for third-party served resources, and methods to request these resources such that all websites use the same order.
Another option would be to rethink the trend of creating web development resources as a service, and go back to having first parties hosting and serving the resources. 
In this spirit, initiatives such as Web Bundles~\citep{webbundles} (though raising other privacy concerns) would remove the need for coordinating between third parties, simplifying the implementation of defenses, and removing vantage points.

\para{IPs \& Anonymity set sizes.}
In the QUIC setting, the adversary is largely helped by the IPs addresses; they can be used to turn the website fingerprinting problem into a closed-world classification problem, to dissect traffic based on first and third parties, and to link together a client's packets.

To address the easy linking of packets, clients could use techniques such as  MIMIQ~\citep{govil2020mimiq} to leverage QUIC's connection migration feature to change their IP address; or Near-Path NAT~\citep{nearpathnat} or MASQUE~\citep{masque} to completely hide their IPs; or CoMPS~\citep{wang2021comps} to hide IPs and split traffic across multiple paths. 
This would force the adversary to probabilistically stitch packets together to form traces.
If the source/destination IP/port are not identifying one client, simply rotating QUIC's connection ID might also prevent the adversary from linking together one client's packets.

To address the ease of separating traffic, CDNs that also host websites could proxy the traffic to third parties, such that all traffic is served from a single IP.\@
Finally, the closed-world size could be increased by co-hosting multiple websites on one server, or making a larger number of websites available behind load balancers; or even moved to open world if all web traffic would be downloaded via anonymous communication networks (\eg Tor~\citep{dingledine2004tor}) or VPNs.

\bibliographystyle{IEEEtranS}
\bibliography{urls,bibliography}

\begin{thebibliography}{10}
\providecommand{\url}[1]{#1}
\csname url@samestyle\endcsname
\providecommand{\newblock}{\relax}
\providecommand{\bibinfo}[2]{#2}
\providecommand{\BIBentrySTDinterwordspacing}{\spaceskip=0pt\relax}
\providecommand{\BIBentryALTinterwordstretchfactor}{4}
\providecommand{\BIBentryALTinterwordspacing}{\spaceskip=\fontdimen2\font plus
\BIBentryALTinterwordstretchfactor\fontdimen3\font minus
  \fontdimen4\font\relax}
\providecommand{\BIBforeignlanguage}[2]{{%
\expandafter\ifx\csname l@#1\endcsname\relax
\typeout{** WARNING: IEEEtranS.bst: No hyphenation pattern has been}%
\typeout{** loaded for the language `#1'. Using the pattern for}%
\typeout{** the default language instead.}%
\else
\language=\csname l@#1\endcsname
\fi
#2}}
\providecommand{\BIBdecl}{\relax}
\BIBdecl

\bibitem{VPNconsumer}
``{3rd Annual VPN Market Report},''
  \url{https://www.security.org/resources/vpn-consumer-report-annual/},
  accessed: 2022-11-29.

\bibitem{alexa1M}
``{Alexa 1M},'' \url{http://s3.amazonaws.com/alexa-static/top-1m.csv.zip},
  accessed: 2021-07-05.

\bibitem{blanket}
``{BLANKET},'' \url{https://github.com/SPIN-UMass/BLANKET}, accessed:
  2022-05-25.

\bibitem{cdnstats}
``{CDN Usage Distribution in the Top 1 Million Sites},''
  \url{https://trends.builtwith.com/cdn}, accessed: 2022-12-02.

\bibitem{umbrella1m}
``{Cisco Umbrella Top 1M Domains List},''
  \url{https://www.trisul.org/devzone/doku.php/cisco_umbrella_top-1m_domains_list},
  accessed: 2021-07-05.

\bibitem{doh}
``{DNS Queries over HTTPS (DoH)},''
  \url{https://datatracker.ietf.org/doc/html/rfc8484}, accessed: 2021-07-05.

\bibitem{tracker-radar}
``{DuckDuckGo Tracker Radar},''
  \url{https://github.com/duckduckgo/tracker-radar}, accessed: 2021-07-05.

\bibitem{masque}
``{Multiplexed Application Substrate over QUIC Encryption (MASQUE)},''
  \url{https://datatracker.ietf.org/wg/masque/about/}, accessed: 2021-10-08.

\bibitem{nearpathnat}
``{Near-path NAT for IP Privacy},''
  \url{https://github.com/bslassey/ip-blindness/blob/master/near_path_nat.md},
  accessed: 2021-10-08.

\bibitem{openwpm}
``{OpenWPM: a Web Privacy Measurement Framework},''
  \url{https://github.com/mozilla/OpenWPM}, accessed: 2021-10-04.

\bibitem{quic_rfc}
``{RFC 9000},'' \url{https://datatracker.ietf.org/doc/html/rfc9000}, accessed:
  2021-08-12.

\bibitem{quic_padding_frames}
``{RFC 9000, Section 19.1 PADDING Frames},''
  \url{https://datatracker.ietf.org/doc/html/rfc9000#section-19.1}, accessed:
  2021-08-12.

\bibitem{dot}
``{Specification for DNS over Transport Layer Security (TLS)},''
  \url{https://datatracker.ietf.org/doc/html/rfc7858}, accessed: 2021-07-05.

\bibitem{majestic1M}
``{The Majestic Million},''
  \url{https://majestic.com/reports/majestic-million}, accessed: 2021-07-05.

\bibitem{ech}
``{TLS Encrypted Client Hello},''
  \url{https://datatracker.ietf.org/doc/html/draft-ietf-tls-esni-12}, accessed:
  2021-07-05.

\bibitem{quic_adoption}
``{Usage statistics of HTTP/3 for websites},''
  \url{https://w3techs.com/technologies/details/ce-http3}, accessed:
  2022-05-27.

\bibitem{webbundles}
``{Web Bundles},''
  \url{https://wicg.github.io/webpackage/draft-yasskin-wpack-bundled-exchanges.html},
  accessed: 2021-10-08.

\bibitem{bai2020fastfe}
J.~Bai, M.~Zhang, G.~Li, C.~Liu, M.~Xu, and H.~Hu, ``{FastFE: Accelerating
  ml-based traffic analysis with programmable switches},'' in \emph{Proceedings
  of the Workshop on Secure Programmable Network Infrastructure}, 2020, pp.
  1--7.

\bibitem{barradas2021flowlens}
D.~Barradas, N.~Santos, L.~Rodrigues, S.~Signorello, F.~M. Ramos, and
  A.~Madeira, ``{FlowLens: Enabling Efficient Flow Classification for ML-based
  Network Security Applications},'' in \emph{Proceedings of the 28th Network
  and Distributed System Security Symposium (San Diego, CA, USA}, 2021.

\bibitem{bhat2018var}
S.~Bhat, D.~Lu, A.~Kwon, and S.~Devadas, ``{Var-CNN: A data-efficient website
  fingerprinting attack based on deep learning},'' \emph{PETS}, 2019.

\bibitem{cai2014cs}
X.~Cai, R.~Nithyanand, and R.~Johnson, ``{CS-BuFLO: A congestion sensitive
  website fingerprinting defense},'' in \emph{Proceedings of the 13th Workshop
  on Privacy in the Electronic Society}, 2014, pp. 121--130.

\bibitem{cai2014tamaraw}
X.~Cai, R.~Nithyanand, T.~Wang, R.~Johnson, and I.~Goldberg, ``{A systematic
  approach to developing and evaluating website fingerprinting defenses},'' in
  \emph{Proceedings of the 2014 ACM SIGSAC Conference on Computer and
  Communications Security}, 2014, pp. 227--238.

\bibitem{cai2012touching}
X.~Cai, X.~C. Zhang, B.~Joshi, and R.~Johnson, ``{Touching from a distance:
  Website fingerprinting attacks and defenses},'' in \emph{Proceedings of the
  2012 ACM conference on Computer and communications security}, 2012, pp.
  605--616.

\bibitem{carela2011analysis}
V.~Carela-Espa{\~n}ol, P.~Barlet-Ros, A.~Cabellos-Aparicio, and
  J.~Sol{\'e}-Pareta, ``{Analysis of the impact of sampling on NetFlow traffic
  classification},'' \emph{Computer Networks}, vol.~55, no.~5, pp. 1083--1099,
  2011.

\bibitem{cheng1998traffic}
H.~Cheng and R.~Avnur, ``{Traffic analysis of SSL encrypted web browsing},''
  \emph{Project paper, University of Berkeley}, 1998.

\bibitem{cherubin2017website}
G.~Cherubin, J.~Hayes, and M.~Ju{\'a}rez, ``{Website Fingerprinting Defenses at
  the Application Layer.}'' \emph{Proc. Priv. Enhancing Technol.}, vol. 2017,
  no.~2, pp. 186--203, 2017.

\bibitem{davidson2022tango}
A.~Davidson, M.~Frei, M.~Gartner, H.~Haddadi, A.~Perrig, J.~S. Nieto,
  P.~Winter, and F.~Wirz, ``Tango or square dance? how tightly should we
  integrate network functionality in browsers?'' in \emph{Proceedings of the
  21st ACM Workshop on Hot Topics in Networks}, 2022, pp. 205--212.

\bibitem{dingledine2004tor}
R.~Dingledine, N.~Mathewson, and P.~Syverson, ``{Tor: The second-generation
  onion router},'' Naval Research Lab Washington DC, Tech. Rep., 2004.

\bibitem{dyer2012peek}
K.~P. Dyer, S.~E. Coull, T.~Ristenpart, and T.~Shrimpton, ``{Peek-a-boo, i
  still see you: Why efficient traffic analysis countermeasures fail},'' in
  \emph{2012 IEEE symposium on security and privacy}.\hskip 1em plus 0.5em
  minus 0.4em\relax IEEE, 2012, pp. 332--346.

\bibitem{feamster2004location}
N.~Feamster and R.~Dingledine, ``{Location diversity in anonymity networks},''
  in \emph{Proceedings of the 2004 ACM workshop on Privacy in the electronic
  society}, 2004, pp. 66--76.

\bibitem{gong2020zero}
J.~Gong and T.~Wang, ``{Zero-delay lightweight defenses against website
  fingerprinting},'' in \emph{{29th USENIX Security Symposium (USENIX Security
  20)}}, 2020, pp. 717--734.

\bibitem{govil2020mimiq}
Y.~Govil, L.~Wang, and J.~Rexford, ``{MIMIQ: Masking IPs with Migration in
  QUIC},'' in \emph{{10th USENIX Workshop on Free and Open Communications on
  the Internet (FOCI 20)}}, 2020.

\bibitem{hayes2016kfingerprinting}
J.~Hayes and G.~Danezis, ``{k-fingerprinting: A Robust Scalable Website
  Fingerprinting Technique},'' in \emph{25th {USENIX} Security Symposium
  ({USENIX} Security 16)}.\hskip 1em plus 0.5em minus 0.4em\relax Austin, TX:
  {USENIX} Association, Aug. 2016, pp. 1187--1203.

\bibitem{hentgen2019measuring}
E.~Hentgen, ``{Measuring the Security of Website Fingerprinting Defenses},''
  \url{https://infoscience.epfl.ch/record/289258?&ln=en}, 2019, accessed:
  2021-10-08.

\bibitem{herrmann2009website}
D.~Herrmann, R.~Wendolsky, and H.~Federrath, ``{Website fingerprinting:
  attacking popular privacy enhancing technologies with the multinomial
  na{\"\i}ve-bayes classifier},'' in \emph{Proceedings of the 2009 ACM workshop
  on Cloud computing security}, 2009, pp. 31--42.

\bibitem{hintz2002fingerprinting}
A.~Hintz, ``{Fingerprinting websites using traffic analysis},'' in
  \emph{International workshop on privacy enhancing technologies}.\hskip 1em
  plus 0.5em minus 0.4em\relax Springer, 2002, pp. 171--178.

\bibitem{hoang2021ipwf}
N.~P. Hoang, A.~A. Niaki, P.~Gill, and M.~Polychronakis, ``{Domain Name
  Encryption Is Not Enough: Privacy Leakage via IP-based Website
  Fingerprinting},'' in \emph{Proceedings of the 21st Privacy Enhancing
  Technologies Symposium}, ser. {PoPETs '21}, 2021.

\bibitem{izhikevich2022zdns}
L.~Izhikevich, G.~Akiwate, B.~Berger, S.~Drakontaidis, A.~Ascheman, P.~Pearce,
  D.~Adrian, and Z.~Durumeric, ``Zdns: a fast dns toolkit for internet
  measurement,'' in \emph{Proceedings of the 22nd ACM Internet Measurement
  Conference}, 2022, pp. 33--43.

\bibitem{johnson2013users}
A.~Johnson, C.~Wacek, R.~Jansen, M.~Sherr, and P.~Syverson, ``{Users get
  routed: Traffic correlation on Tor by realistic adversaries},'' in
  \emph{Proceedings of the 2013 ACM SIGSAC conference on Computer \&
  communications security}, 2013, pp. 337--348.

\bibitem{juarez2016toward}
M.~Juarez, M.~Imani, M.~Perry, C.~Diaz, and M.~Wright, ``{Toward an efficient
  website fingerprinting defense},'' in \emph{European Symposium on Research in
  Computer Security}.\hskip 1em plus 0.5em minus 0.4em\relax Springer, 2016,
  pp. 27--46.

\bibitem{juen2015defending}
J.~Juen, A.~Johnson, A.~Das, N.~Borisov, and M.~Caesar, ``{Defending Tor from
  Network Adversaries: A Case Study of Network Path Prediction},''
  \emph{Proceedings on Privacy Enhancing Technologies}, vol. 2015, no.~2, pp.
  171--187, 2015.

\bibitem{krishnamurthy2003sketch}
B.~Krishnamurthy, S.~Sen, Y.~Zhang, and Y.~Chen, ``{Sketch-based change
  detection: Methods, evaluation, and applications},'' in \emph{Proceedings of
  the 3rd ACM SIGCOMM conference on Internet measurement}, 2003, pp. 234--247.

\bibitem{liu2016one}
Z.~Liu, A.~Manousis, G.~Vorsanger, V.~Sekar, and V.~Braverman, ``{One sketch to
  rule them all: Rethinking network flow monitoring with univmon},'' in
  \emph{Proceedings of the 2016 ACM SIGCOMM Conference}, 2016, pp. 101--114.

\bibitem{luo2011httpos}
X.~Luo, P.~Zhou, E.~W. Chan, W.~Lee, R.~K. Chang, R.~Perdisci \emph{et~al.},
  ``{HTTPOS: Sealing Information Leaks with Browser-side Obfuscation of
  Encrypted Flows.}'' in \emph{NDSS}, 2011.

\bibitem{perry22rp}
{Mike Perry}, ``{Experimental Defense for Website Traffic Fingerprinting},''
  \url{https://blog.torproject.org/
  blog/experimental-defense-website-traffic-fingerprinting}, accessed:
  2021-10-03.

\bibitem{miller2014know}
B.~Miller, L.~Huang, A.~D. Joseph, and J.~D. Tygar, ``{I know why you went to
  the clinic: Risks and realization of https traffic analysis},'' in
  \emph{International Symposium on Privacy Enhancing Technologies
  Symposium}.\hskip 1em plus 0.5em minus 0.4em\relax Springer, 2014, pp.
  143--163.

\bibitem{murdoch2007sampled}
S.~J. Murdoch and P.~Zieli{\'n}ski, ``{Sampled traffic analysis by
  internet-exchange-level adversaries},'' in \emph{International workshop on
  privacy enhancing technologies}.\hskip 1em plus 0.5em minus 0.4em\relax
  Springer, 2007, pp. 167--183.

\bibitem{nasr2021blind}
M.~Nasr, A.~Bahramali, and A.~Houmansadr, ``{Defeating DNN-Based Traffic
  Analysis Systems in Real-Time With Blind Adversarial Perturbations},'' in
  \emph{{30th USENIX Security Symposium (USENIX Security 21)}}, 2021.

\bibitem{nasr2017compressive}
M.~Nasr, A.~Houmansadr, and A.~Mazumdar, ``{Compressive traffic analysis: A new
  paradigm for scalable traffic analysis},'' in \emph{Proceedings of the 2017
  ACM SIGSAC Conference on Computer and Communications Security}, 2017, pp.
  2053--2069.

\bibitem{nomikos2016traixroute}
G.~Nomikos and X.~Dimitropoulos, ``{traIXroute: Detecting IXPs in traceroute
  paths},'' in \emph{International Conference on Passive and Active Network
  Measurement}.\hskip 1em plus 0.5em minus 0.4em\relax Springer, 2016, pp.
  346--358.

\bibitem{OverdorfJAGD17}
R.~Overdorf, M.~Juarez, G.~Acar, R.~Greenstadt, and C.~Diaz, ``How unique is
  your. onion? an analysis of the fingerprintability of tor onion services,''
  in \emph{Proceedings of the 2017 ACM SIGSAC Conference on Computer and
  Communications Security}, 2017, pp. 2021--2036.

\bibitem{panchenko2016website}
A.~Panchenko, F.~Lanze, J.~Pennekamp, T.~Engel, A.~Zinnen, M.~Henze, and
  K.~Wehrle, ``{Website Fingerprinting at Internet Scale.}'' in \emph{NDSS},
  2016.

\bibitem{patil2019can}
S.~Patil and N.~Borisov, ``What can you learn from an {IP?}'' in
  \emph{Proceedings of the Applied Networking Research Workshop}, 2019, pp.
  45--51.

\bibitem{rahman2020mockingbird}
M.~S. Rahman, M.~Imani, N.~Mathews, and M.~Wright, ``{Mockingbird: Defending
  against deep-learning-based website fingerprinting attacks with adversarial
  traces},'' \emph{IEEE Transactions on Information Forensics and Security},
  vol.~16, pp. 1594--1609, 2020.

\bibitem{rimmer2018automated}
V.~Rimmer, D.~Preuveneers, M.~Juarez, T.~Van~Goethem, and W.~Joosen,
  ``{Automated Website Fingerprinting through Deep Learning},'' in \emph{NDSS},
  2018.

\bibitem{shan2021dolos}
S.~Shan, A.~N. Bhagoji, H.~Zheng, and B.~Y. Zhao, ``{A Real-time Defense
  against Website Fingerprinting Attacks},'' in \emph{ACM Workshop on
  Artificial Intelligence and Security (AISec'21)}, 2021.

\bibitem{siby2019encrypted}
S.~Siby, M.~Juarez, C.~Diaz, N.~Vallina-Rodriguez, and C.~Troncoso,
  ``{Encrypted DNS--> Privacy? A traffic analysis perspective},'' in
  \emph{NDSS}, 2020.

\bibitem{silva2017inside}
J.~M.~C. Silva, P.~Carvalho, and S.~R. Lima, ``{Inside packet sampling
  techniques: exploring modularity to enhance network measurements},''
  \emph{International Journal of Communication Systems}, vol.~30, no.~6, p.
  e3135, 2017.

\bibitem{singanamalla2020oblivious}
S.~Singanamalla, S.~Chunhapanya, M.~Vavru{\v{s}}a, T.~Verma, P.~Wu, M.~Fayed,
  K.~Heimerl, N.~Sullivan, and C.~Wood, ``{Oblivious DNS over HTTPS (ODoH): A
  Practical Privacy Enhancement to DNS},'' \emph{PETS}, 2021.

\bibitem{sirinam2018deep}
P.~Sirinam, M.~Imani, M.~Juarez, and M.~Wright, ``{Deep fingerprinting:
  Undermining website fingerprinting defenses with deep learning},'' in
  \emph{Proceedings of the 2018 ACM SIGSAC Conference on Computer and
  Communications Security}, 2018, pp. 1928--1943.

\bibitem{smith2022qcsd}
J.-P. Smith, L.~Dolfi, P.~Mittal, and A.~Perrig, ``$\{$QCSD$\}$: A
  $\{$QUIC$\}$$\{$Client-Side$\}$$\{$Website-Fingerprinting$\}$ defence
  framework,'' in \emph{31st USENIX Security Symposium (USENIX Security 22)},
  2022, pp. 771--789.

\bibitem{smith2021website}
J.-P. Smith, P.~Mittal, and A.~Perrig, ``{Website Fingerprinting in the Age of
  QUIC.}'' \emph{PETS}, vol. 2021, no.~2, pp. 48--69, 2021.

\bibitem{tammaro2012exploiting}
D.~Tammaro, S.~Valenti, D.~Rossi, and A.~Pescap{\'e}, ``{Exploiting
  packet-sampling measurements for traffic characterization and
  classification},'' \emph{International Journal of Network Management},
  vol.~22, no.~6, pp. 451--476, 2012.

\bibitem{tune2014ofss}
P.~Tune and D.~Veitch, ``{OFSS: Skampling for the flow size distribution},'' in
  \emph{Proceedings of the 2014 Conference on Internet Measurement Conference},
  2014, pp. 235--240.

\bibitem{wang2021comps}
M.~Wang, A.~Kulshrestha, L.~Wang, and P.~Mittal, ``Leveraging strategic
  connection migration-powered traffic splitting for privacy,''
  \emph{Proceedings on Privacy Enhancing Technologies}, vol.~1, p.~18, 2022.

\bibitem{wang2014effective}
T.~Wang, X.~Cai, R.~Nithyanand, R.~Johnson, and I.~Goldberg, ``{Effective
  attacks and provable defenses for website fingerprinting},'' in \emph{{23rd
  USENIX Security Symposium (USENIX Security 14)}}, 2014, pp. 143--157.

\end{thebibliography}

\appendix
\subsection{Traceroute experiments at additional vantage points.}
\label{app:vantage}

The client location impacts the resources that might be fetched during a page load, and the paths taken by the network traffic to the destination servers.
This, in turn, impacts the ASes that can view the traffic. 
In order to confirm that the trends we observe in our traffic visibility experiment (Section~\ref{sec:adv-partial-view}) hold at different locations, we collect additional traceroutes from three additional vantage  points located in Germany, UK, and Singapore.

\begin{figure}[!htpb]
	\begin{subfigure}{\linewidth}
	\includegraphics[width=\textwidth]{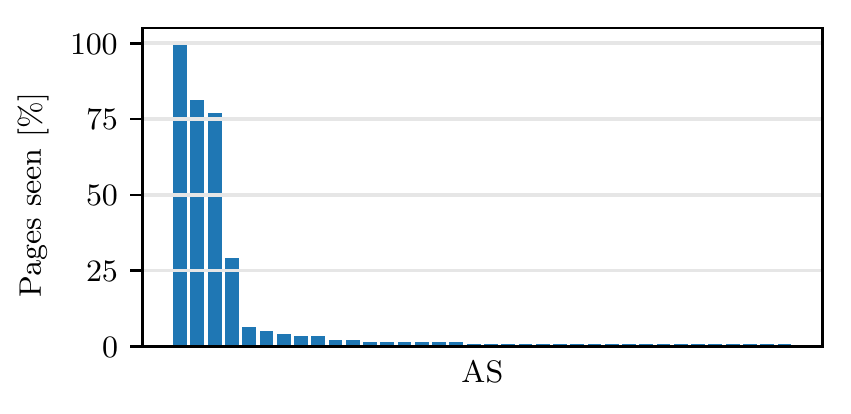}
	\vspace{-2em}
	\caption{\label{fig:quic2} Germany}
	\end{subfigure}
	\begin{subfigure}{\linewidth}
	\includegraphics[width=\textwidth]{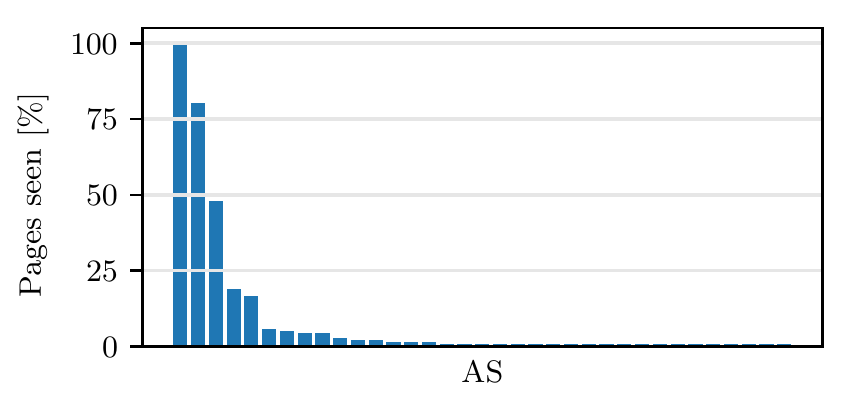}
	\vspace{-2em}
	\caption{\label{fig:quic3} United Kingdom}
	\end{subfigure}
	\begin{subfigure}{\linewidth}
	\includegraphics[width=\textwidth]{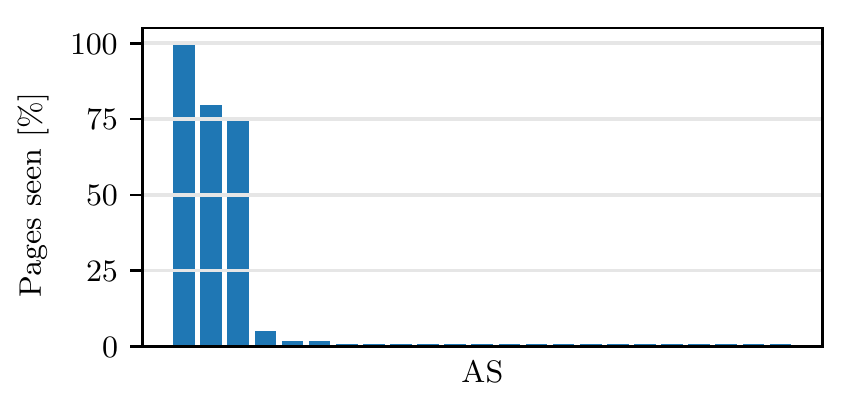}
	\vspace{-2em}
	\caption{\label{fig:quic4} Singapore}
	\end{subfigure}
	\vspace{-0.2em}
	\caption{Distribution of webpages seen by each AS, at three vantage points. Only the client's AS, Google, and Cloudflare observe $>25\%$ of the traffic.}
	\label{fig:vantage}
\end{figure}

Figure~\ref{fig:vantage} shows the distribution of webpages seen by different ASes, for our three vantage points.
The number of total ASes we encounter on the traceroute are 36, 35, and 23.
Out of these 13 ASes are common across all the vantage points.
While the ASes that observe the traffic vary across locations, similar to Section~\ref{sec:adv-partial-view}
,  only a small proportion of ASes that observe a large proportion of the traffic. Three ASes see more than 25\% of the traffic for each vantage point: the client's AS, Google, and Cloudflare.

\subsection{IP anonymity sets}

\begin{figure}[!htpb]
	\centering
	\small
	\includegraphics[width=\linewidth]{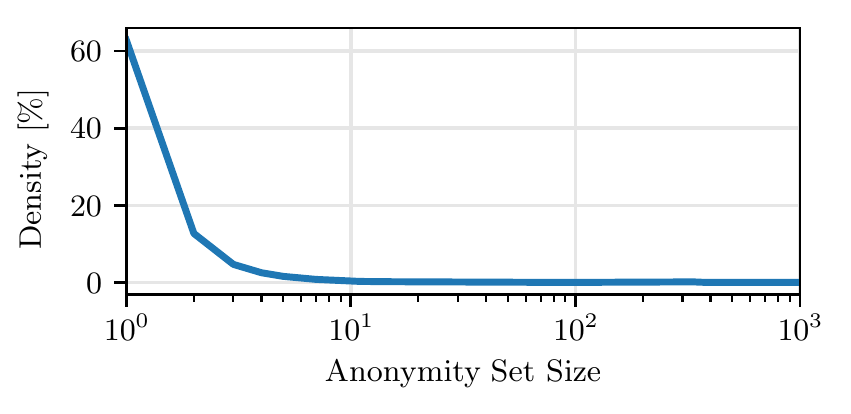}
	\vspace{-0.6cm}
	\caption{Distribution of the cluster sizes of $1.3$M domains.}
	\label{fig:distribution-clusters}
	\vspace{-0.3cm}
\end{figure}

Figure~\ref{fig:distribution-clusters} shows the distribution of clusters for the 1.3M domains we obtained from our CDN partner. We find that 60\% of the clusters have an anonymity set of one, i.e., the domains are hosted on a unique IP. These domains can be identified without conducting website fingerprinting. Only 8.5\% of the IPs host more than 150 domains.

\subsection{Additional feature importance graphics}
\label{app:feature_importance}

\begin{figure}[!htpb]
	\centering
	\includegraphics[width=\linewidth]{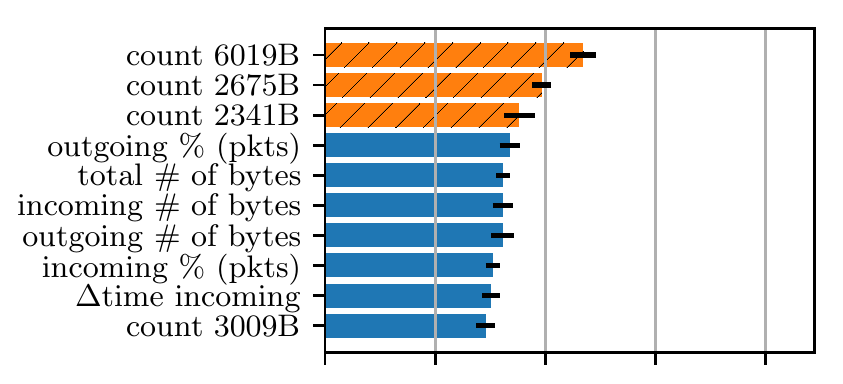}
	\caption{Feature importance for the cluster dataset. Due to low prevalence of QUIC, most of the features are TLS-specific (orange, dashed).}
	\label{fig:mixed-fi}
\end{figure}

Figure~\ref{fig:mixed-fi} shows the feature importance when the classifier is run on the dataset from the cluster. We get an F1-score of 66.6\% (std. dev. 0.5). On running a feature analysis, we find that the most important features are TLS-specific. 
Since these features cannot be protected with a QUIC PADDING frame, we discard this dataset in favor of QUIC-dominated datasets created from website lists.

\begin{figure}[!htpb]
	\centering
	\includegraphics[width=\linewidth]{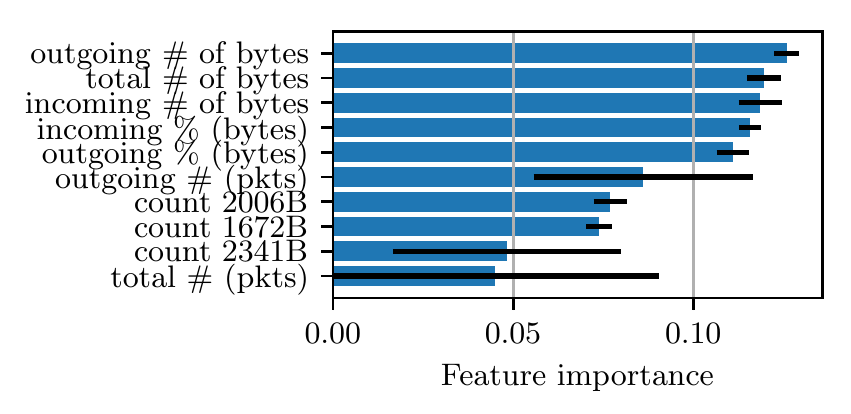}
	\caption{Feature importance for \quicM when using application-layer features (based on HAR captures).}
	\label{fig:har-quic150-fi}
\end{figure}

Figure~\ref{fig:har-quic150-fi} shows the feature importance of \quicM when using application layer features from HAR captures. We find that size based features are the most important, and time-based features do not play a large role.

\begin{figure}[!htpb]
	\centering
	\includegraphics[width=\linewidth]{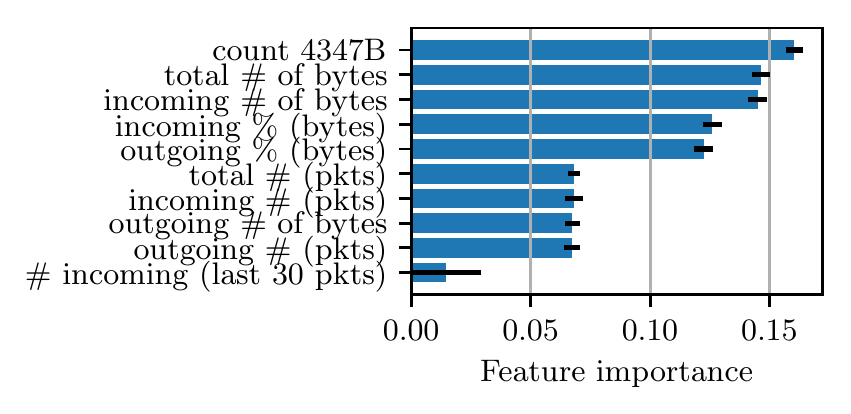}
	\caption{Feature importance with $3$ padding sizes: $5.58$ kB, $21$ kB, $3.6$ MB.}
	\label{fig:har-quic150-pad-3steps-fi}
\end{figure}

Figure~\ref{fig:har-quic150-pad-3steps-fi} shows the feature importance when we protect local features at the application layer with a padding function.
The function uses a parameter, $N$, that indicates the number of sizes to which resources can be padded. 
We experiment with various values of $N$, and find that only large amounts of padding (small $N$) have an effect on the adversary's performance. 
This is because the adversary still has access to the global features such as number of requests and overall volume (as shown by Figure~\ref{fig:har-quic150-pad-3steps-fi} for an example of $N = 3$.

\clearpage
\onecolumn
\MarkupsHowto 

\end{document}